\documentclass[aps,pra,twocolumn,floatfix,superscriptaddress]{revtex4-1}
\usepackage{subfigure,dcolumn}
\usepackage[T2A,T1]{fontenc}
\usepackage[english]{babel}
\usepackage{braket}
\usepackage{graphicx}
\usepackage[colorinlistoftodos]{todonotes}
\usepackage[utf8]{inputenc}
\usepackage{amsthm}
\usepackage{amsmath}
\usepackage{amsfonts}
\usepackage{slashed}
\usepackage{hyperref}
\usepackage[normalem]{ulem}
\usepackage{soul}
\usepackage{MnSymbol}

\hypersetup{colorlinks,linkcolor=red,citecolor=blue,urlcolor=blue}

\begin{document}

\title{Theory of waveguide-QED with moving emitters}

\author{Eduardo Sánchez-Burillo}
\affiliation{Max-Planck-Institut für Quantenoptik, Hans-Kopfermann-Str. 1, 85748 Garching, Germany}

\author{Alejandro González-Tudela}
\affiliation{Instituto de Física Fundamental IFF-CSIC, Calle Serrano 113b, E-28006 Madrid, Spain}

\author{Carlos Gonzalez-Ballestero}
\email{carlos.gonzalez-ballestero@uibk.ac.at}
\affiliation{Institute for Quantum Optics and Quantum Information of the Austrian Academy of Sciences, A-6020 Innsbruck, Austria}
\affiliation{Institute for Theoretical Physics, University of Innsbruck, A-6020 Innsbruck, Austria}

\date{\today}

\begin{abstract}
We theoretically study a system composed by a waveguide and a quantum emitter moving in a one-dimensional potential along the waveguide axis. We focus on the single excitation subspace and treat the emitter motional degree of freedom full quantum mechanically. 
We first characterize single-photon scattering off a single moving quantum emitter, showing both direction-dependent transmission and recoil-induced reduction of the quantum emitter motional energy. We then characterize the bound states within the bandgap, which display a motion-induced asymmetric phase in real space. We also demonstrate how these bound states form a continuous band with exotic dispersion relations. Finally, we study the spontaneous emission of an initially excited quantum emitter with various initial momentum distributions, finding strong deviations with respect to the static emitter counterpart both in the occupation dynamics and in the spatial distribution of the emitted photons. Our work extends the waveguide-QED toolbox by including the quantum motional degree of freedom of emitters, whose impact in the few-photon dynamics could be harnessed  for quantum technologies.
\end{abstract}

%\pacs{42.50.Ct, 42.50.-p, 03.65.-w, 11.55.Bq}

%42.50.Ex, 42.50.Pq

\maketitle

%\tableofcontents
%%%%%%%%%%%%%%%%%%%%%%%%%%%%%%%%%%%%%%%%%%
%%%%%%%%%%%%%%%%%%%%%%%%%%%%%%%%%%%%%%%%%%
%%%%%%%%%%%%%%%%%%%%%%%%%%%%%%%%%%%%%%%%%%
%%%%%%%%%%%%%%%%%%%%%%%%%%%%%%%%%%%%%%%%%%

\section{Introduction}

The possibility of coupling quantum emitters to engineered reservoirs, for instance photonic nanostructures, lies at the basis of modern quantum technologies. One-dimensional reservoirs are especially appealing in this regard since they combine enhanced light-emitter couplings with open propagation directions, making them ideal quantum connectors within quantum networks \cite{KimbleNature2008}. The resulting field of waveguide-QED (wQED) has experienced a large development in the last decade, both theoretically and experimentally, with the demonstration of, among others, few-photon nonlinear devices \cite{ChangNatPhys2007}, cascaded quantum systems based on chirality \cite{LodahlNature2017}, or bound light-matter states within a bandgap \cite{KrinnerNature2018} resulting in tuneable qubit-qubit interactions \cite{Hood2016}. In most experimental works, the motional degrees of freedom of quantum emitters have received little attention, except for the demonstration of motional cooling of quantum emitters coupled to waveguides~\cite{YuAPL2014,GobanPRL2015,MengPRX2018}. This is in sharp contrast to other related fields such as e.g. three-dimensional optical lattices \cite{BlochZollerChapter,GreinerNature2008}, where emitter motion has not only been analyzed but harnessed for generating non-classical states \cite{CiracAcPressChapter}, engineering qubit-qubit interactions \cite{MunstermannPRL2000,AsbothPRA2004}, trapping emitters with single photons \cite{HoodScience2000,PinkseNature2000}, or devising quantum protocols in the ultrastrong coupling regime~\cite{DareauPRL2018}.

\begin{figure}
    \centering
    \includegraphics[width=0.8\linewidth]{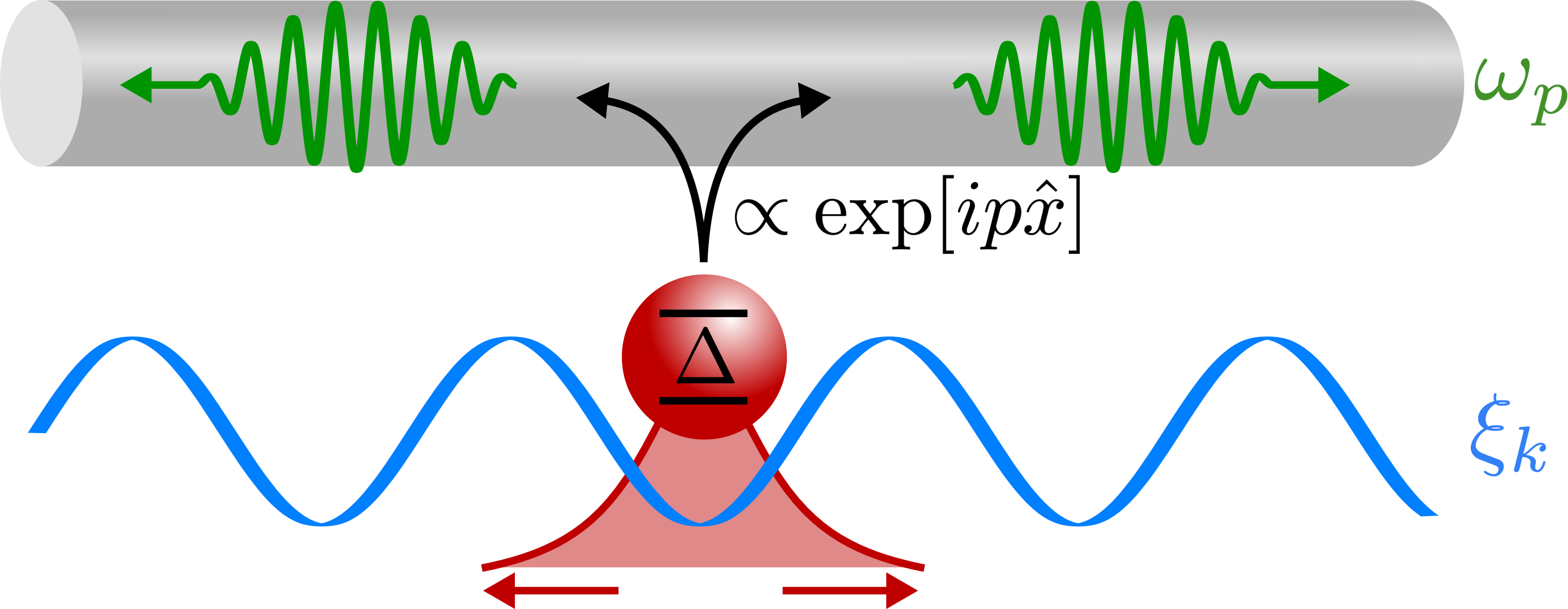}
    \caption{Sketch of the system under study. A qubit with frequency $\Delta$ can move in a given potential with dispersion $\xi_k$, $k$ being the qubit momentum. The qubit interacts through electric dipole interaction with the photons in a waveguide with dispersion relation $\omega_p$, $p$ being the photon wavevector. The nonlinear coupling between photons of momentum $p$ and the internal resonance of the qubit is proportional to the photonic field amplitude at the qubit position, $\exp[ip\hat{x}]$.}
    \label{fig:system}
\end{figure}

The theoretical description of the motional degrees of freedom of quantum emitters in waveguide QED, as well as of the possibility of harnessing them, has only recently been explored by pioneering works \cite{Calajo2017}. These works address the important limit of quantum emitters moving along classical trajectories, where the motional degrees of freedom can be treated classically. Such a high-kinetic-energy regime applies to emitters with much larger momentum than the photons, for instance emitters in room-temperature thermal motional states. 
However, the theoretical description of the opposite limit, namely low kinetic energy of the quantum emitters, is lacking. Deriving this description is important as 
such regime has become attainable in many waveguide QED systems, for instance low-mass emitters motionally cooled near the ground state in waveguide optical lattices~\cite{YuAPL2014,GobanPRL2015,MengPRX2018}. At low kinetic energies, the motion of the quantum emitters must be treated quantum mechanically, as their linear momentum can become comparable both to their de Broglie momentum and to the momentum of single optical photons \cite{LeonardiOptComm1993,MortezapourLPL2017}. Furthermore, a full quantum description of the emitter motion could result in new capabilities to harness, as recently demonstrated for atoms near parabolic mirrors \cite{TrautmannPRA2016}.  Motivated by the above results, in this paper we theoretically study the properties of a single-qubit waveguide QED system in the single excitation subspace by fully including the quantum motional degrees of freedom of the quantum emitter.

This paper is organized as follows. In Sec. \ref{SecHamiltonian}, we describe the wQED Hamiltonian in the presence of qubit motion and the conserved quantities of the system. We also discuss physically motivated expressions for both the qubit- and the waveguide dispersion relations. In Sec.~\ref{SecScattering}, we compute the single-photon S-matrix and the scattering coefficients. In Sec.~\ref{SecBoundStates}, we study the bound states appearing within the bandgap, as well as their existence conditions, their energies, and their position-space representation. In Sec.~\ref{sect:se}, we study the spontaneous emission of an excited qubit in different initial states, characterizing the different emission regimes and computing the spatial distribution of both the qubit and the photonic wavefunctions. Finally, our conclusions are presented in Sec.~\ref{SecConclusions}.

\section{System and Hamiltonian}\label{SecHamiltonian}

\subsection{Hamiltonian and constants of motion}\label{SubsecHgeneral}

The system under consideration, schematically depicted in Fig.~\ref{fig:system}, consists on a quantum emitter trapped in an external potential and coupled to a waveguide, which for simplicity we consider single-band. The emitter is described by both motional and internal (e.g. electronic) degrees of freedom. The internal degrees of freedom are modelled as a two-level system with ground and excited states $\vert g\rangle$ and $\vert e\rangle$, whose energy splitting is $\Delta$. On the other hand, the motional state of the qubit is described by its momentum $k$. The Hamiltonian of the system can be written as
\begin{equation}\label{eq:H}
    H = H_\text{bare} + H_\text{int},
\end{equation}
where $H_\text{bare}$ is the free Hamiltonian of qubit and waveguide and $H_\text{int}$ the qubit-photon interaction term. The former is given by ($\hbar=1$)
\begin{equation}\label{eq:Hb}
    H_{\rm bare} = \sum_p \omega_p a_p^\dagger a_p + \sum_k \xi_k\left(e_k^\dagger e_k + g_k^\dagger g_k\right) + 
    \sum_k\Delta e_k^\dagger e_k.
\end{equation}
Here, the first term corresponds to the Hamiltonian of the waveguide of photons, with wavevector $p$, single-band dispersion relation $\omega_p$ and bosonic ladder operators $a_p$ and $a_p^\dagger$. 
The second and third terms represent the energy of the qubit, expressed for convenience in terms of the compound motional+electronic operators $e_k^\dagger$ and $g_k^\dagger$, which correspond to creation of a qubit state with momentum $k$ and internal state $\vert e\rangle$ and $\vert g \rangle$, respectively. Formally, these operators can be defined as $e_k^\dagger\equiv e^\dagger c^\dagger_k$ and $g_k^\dagger\equiv g^\dagger c^\dagger_k$. Here, the operators  $c^\dagger_k$ create a qubit motional state with momentum $k$, and obey bosonic/fermionic statistics depending on the bosonic/fermionic character of the qubit. Conversely, the operators $e^\dagger$ and $g^\dagger$ are defined through their actions on the vacuum, namely $e^\dagger\vert 0\rangle = \vert e\rangle$ and $g^\dagger\vert 0\rangle = \vert g\rangle$ respectively. These operators, which we have defined for notational convenience, are not necessarily physically meaningful. Indeed, any system observable depends only on the \textit{products} of these operators, which have a well-defined physical interpretation in terms of the Pauli matrices for a two-level system, namely $2e^\dagger e=1+\sigma_z$, $2g^\dagger g=1-\sigma_z$, $eg^\dagger=0$, and $e^\dagger g = \sigma^\dagger$, with $\sigma^\dagger\equiv\vert e\rangle\langle g\vert$ and $\sigma_z \equiv  \sigma^\dagger\sigma-\sigma\sigma^\dagger$. 
The dispersion relation for the qubit motion, $\xi_k$ in Eq. \eqref{eq:Hb}, is assumed to be independent of its electronic state. 
The second contribution in Eq.~\eqref{eq:H} contains the interaction between waveguide photons and the qubit degrees of freedom in the electric dipole approximation. In the rotating-wave approximation  \cite{Cohen-Tannoudji1992}, valid when the coupling is small enough compared to the remaining energy scales of the system, this term reads~\footnote{Note that this Hamiltonian is an extension of the Fr{\"o}hlich polaron Hamiltonian to an impurity with two internal levels  \cite{KainPRA2016}}
\begin{equation}\label{eq:Hint}
\begin{split}
   H_\text{int} &=
   \sum_p \Omega_p\left(e^{ipX}a_p\sigma^\dagger+\text{H.c.}\right)
   \\
   &
   =\frac{1}{\sqrt{L}}\sum_{p,k} \Omega_{p}  e_{k+p}^\dagger a_p\, g_k + \text{H.c.}
\end{split}
\end{equation}
Here $L$ is the number of photonic modes and $\Omega_{p}$ the coupling rate between the qubit transition and the photonic mode $p$.  Note that, although for convenience we choose a discrete labeling of the momenta, taking the continuum limit is straightforward as we will see below.

The Hamiltonian Eq.~\eqref{eq:H} 
commutes with the total excitation number
\begin{equation}
\mathcal{N} = \sum_p a_p^\dagger a_p + \sum_k e_k^\dagger e_k, \label{eq:number}
\end{equation}
thus allowing to restrict further analysis to a given Fock subspace. Specifically, we aim at studying and characterizing the single-excitation subspace ($\mathcal{N}=1$). Furthermore, the Hamiltonian also commutes with the total momentum
\begin{equation}
\mathcal{K} = \sum_p p\, a_p^\dagger a_p + \sum_k k(e_k^\dagger e_k + g_k^\dagger g_k).  \label{eq:totalmomentum}  
\end{equation}
This allows to further simplify our study by working within a subspace with a fixed value of the total momentum
 $\mathcal{K}=K$. It is possible to simplify the Hamiltonian by projecting Eq. \eqref{eq:H} into this subspace, obtaining
\begin{align}\label{eq:H1K}
    H_{1,K} & = \sum_p (\omega_p + \xi_{K-p})\ket{p}_K \bra{p}_K + (\Delta + \xi_K)\ket{K}\bra{K} \nonumber \\
    & + \sum_p\frac{\Omega_p}{\sqrt{L}}(\ket{p}_K \bra{K} + \text{H.c.}),
\end{align}
where the sub-indices $1,K$ indicate the eigenvalues of $\mathcal{N}$ and $\mathcal{K}$ respectively, and we have defined the states
\begin{equation}\label{eq:states}
    \begin{split}
        &\ket{p}_K \equiv a_p^\dagger g_{K-p}^\dagger \ket{0}, \\
        &\ket{K} \equiv e_K^\dagger \ket{0},
    \end{split}
\end{equation}
with $\ket{0}$ the vacuum state of $a_p$, $e_k$, and $g_k$.  The projected Hamiltonian Eq. \eqref{eq:H1K} has a similar form as that of a \emph{static} emitter coupled to a waveguide. Specifically, our system in the subspace $\{1,K\}$ is equivalent to a continuum of hybrid photonic+motional states $\{ \ket{p}_K\}$ with effective dispersion relation
\begin{equation}\label{eq:dispersionK}
\tilde{\omega}_{K,p} \equiv \omega_p + \xi_{K-p}    
\end{equation}
coupled to a two-level impurity with modified level splitting
\begin{equation}\label{eq:impurityK}
    E_{K,\Delta} \equiv \Delta + \xi_K.
\end{equation}
In the following, we will use the Hamiltonian in version \eqref{eq:H1K} to derive our results.

\subsection{Case study}\label{SubSeccasestudy}

Although most of the results derived below are general, it will be convenient for the discussion to consider a particular case. Specifically,  we will assume for simplicity a constant coupling rate, namely 
 $\Omega_{p}=\Omega$, and the following forms for the dispersion relations,
\begin{align}
\omega_p & = -2J\cos p, \label{eq:wp} \\
\xi_k & = -2J' \cos k. \label{eq:xik}
\end{align}
These expressions describe both a qubit and a photon confined to a linear array of localized sites, with nearest-neighbour couplings $J$ and $J'$ respectively and lattice spacing equal to 1. In the above expressions and hereafter, all momenta such as $k$ and $p$ will be expressed in units of the inverse lattice spacing and thus represent dimensionless quantities. 
The corresponding group velocities for the photon and qubit are
\begin{align}
    v^\text{ph}_p = & \frac{d\omega_p}{dp} = 2J\sin p, \label{eq:vph} \\
    v^\text{qb}_k = & \frac{d\xi_k}{dk} = 2J'\sin k. \label{eq:vqb}
\end{align}
The specific situation described by Eqs. \eqref{eq:wp} and \eqref{eq:xik} can be implemented in various setups. A paradigmatic example could be optical lattices, either generated by optical waveguides \cite{YuAPL2014,GobanPRL2015,MengPRX2018}, coupled to optical cavities \cite{CaballeroPRA2016} or mechanical modes \cite{VochezerPRL2018}, or tailored such that the roles of qubit motion and photonic modes are played by other degrees of freedom, such as spin states \cite{ForsterPRL2009}, different species of quantum emitters \cite{LeePRA2005,BarbieroSciAdv2019}, or free (untrapped) atomic transitions \cite{deVegaPRL2008,KrinnerNature2018,NavarreteNJP2011}.
An alternative possibility to engineer the above interactions is to use mechanical degrees of freedom, either in the form of acoustic waves modulating optical waveguides 
 \cite{CalajoPRA2019}, or in the form of acoustic lattices for other particles and quasiparticles, such as electrons \cite{SchuetzPRX2017} or exciton-polaritons \cite{CerdaJPD2017,Chafatinos2020}. Implementations in microwave quantum devices could also be engineered using e.g. flying Rydberg atoms \cite{MorganPRA2018}.

A second advantage of the above particularizations, namely of Eqs.~\eqref{eq:wp} and \eqref{eq:xik}, is that the chosen dispersion relations are general enough to allow the Hamiltonian to recover relevant limits. First, the static qubit limit corresponds to $J'\to 0$, namely to a vanishing hopping amplitude between neighboring sites of the qubit lattice. Second, well-known dispersion relations can also be recovered for particular values of $k$ or $p$, provided that the coupling $\Omega$ is weak enough so that the changes in such wavevectors during the time evolution remain small. Specifically,  for $\Omega \ll J$ and $p\approx 0$ ($p\approx\pi/2$) it corresponds to a qubit within a quadratic (linear) photonic band, namely $\omega_p\approx p^2$ ($\omega_p\approx p-\pi/2$).
Moreover, the important case of a free-qubit is recovered for $\Omega \ll J'$ and $k\approx 0$, where the qubit energy becomes quadratic, i.e. $\xi_k \approx -2J' + J'k^2$. In this limit the qubit behaves as a free particle with effective mass $1/2J'$.

\section{Single-photon scattering}\label{SecScattering}

In this section we characterize the scattering of a single photon off a moving emitter initially in its internal ground state. We start in Sec. \ref{SubsecProcesses} by defining and computing the $S-$matrix, and studying the different allowed scattering processes. Then, in Sec. \ref{SubsecAmplitudes}, we compute the probability amplitudes of such scattering processes and demonstrate nonreciprocal single-photon scattering induced by qubit motion and the possibility of reducing the motional energy (cooling) of the qubit through single-photon scattering.

\subsection{$S-$matrix and allowed scattering processes}\label{SubsecProcesses}

Our goal is to compute the asymptotic scattering amplitudes for an initial state characterized by a single photon with momentum $p_i$ and a ground-state qubit with momentum $k_i$. Note that, 
in the spirit of our interpretation of Eq.~\eqref{eq:H1K}, this is equivalent to the scattering of a particle with input energy $\tilde{\omega}_{k_i+p_i,p_i} = \omega_{p_i} + \xi_{k_i}$ interacting locally with a two-level qubit with energy gap $E_{k_i+p_i,\Delta} = \xi_{k_i+p_i}+\Delta$. To characterize the scattering, we compute the scattering matrix
\begin{equation}
S_{fi} \equiv \bra{f}U(+\infty,-\infty)\ket{i},    \label{eq:Sfi}
\end{equation}
with $\ket{i}$ and $\ket{f}$ the initial and final states and $U(+\infty,-\infty)$ the time-evolution operator connecting the distant past and the distant future, associated to the Hamiltonian Eq.~\eqref{eq:H1K} and in the interaction picture with respect to its free part. Since at $t\to\infty$ the qubit will remain in its ground state, we can choose the initial and final states as
\begin{align}
    \ket{i} = & \ket{p_i}_{k_i+p_i}, \label{eq:i} \\
    \ket{f} = & \ket{p_f}_{k_f+p_f}, \label{eq:f}
\end{align}
without loss of generality. The single-photon $S-$matrix describes the probability amplitude of scattering from state $\ket{i}$ to state $\ket{f}$, and is calculated for general dispersion relations in Appendix \ref{app:s_matrix}.

\begin{figure}
    \centering
    \includegraphics[width=0.9\linewidth]{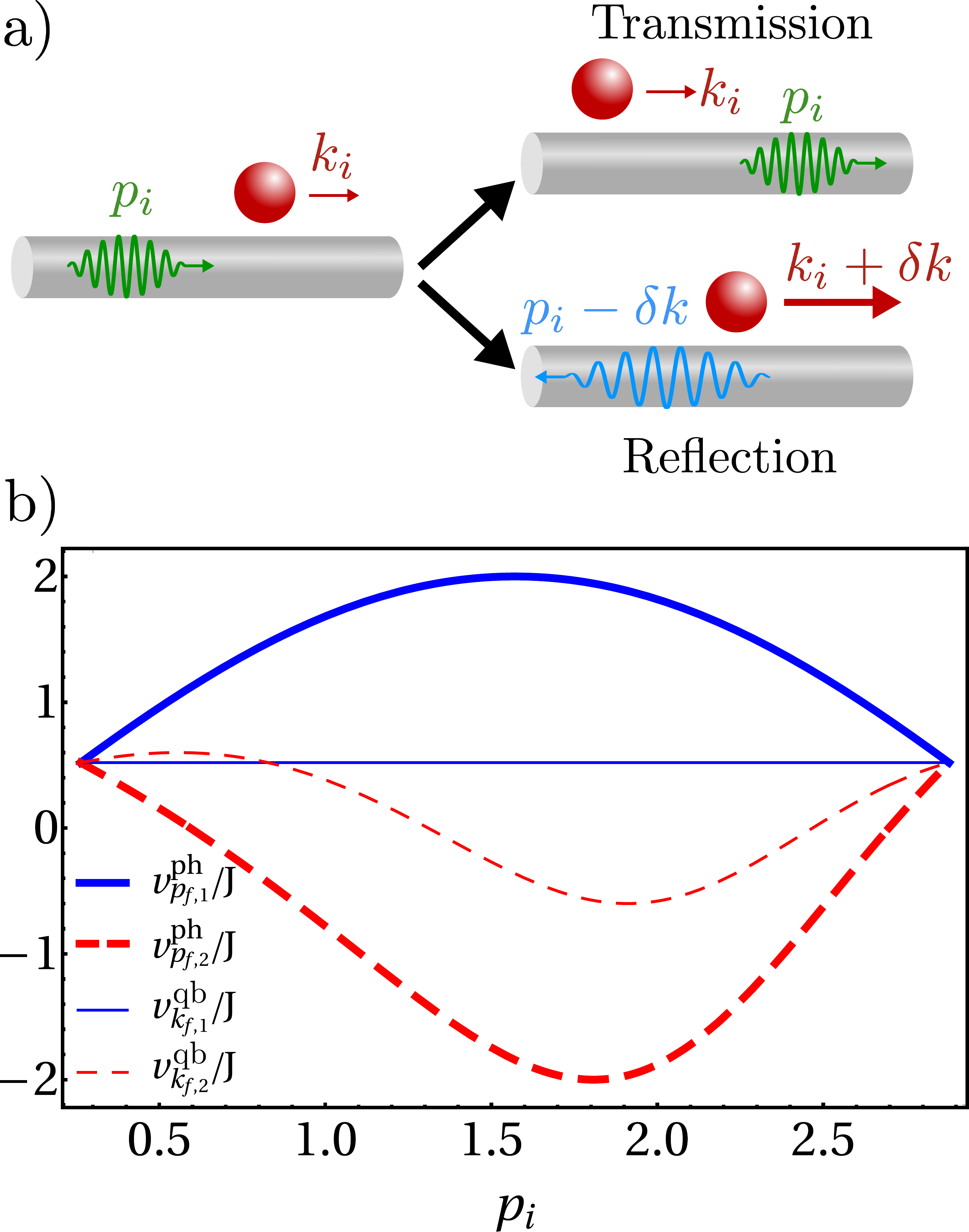}
    \caption{a) Illustration of the two allowed scattering processes, namely elastic transmission and inelastic reflection. b) Final group velocities of photon (thick lines) and qubit (thin lines) for the transmission (blue) and reflection (red) processes. Here we choose $J'=0.3J$, $k_i=\pi/3$, and for simplicity restrict the plot to initial momenta $p_i$ for which the velocity of the photon is larger than that of the qubit, i.e., 
     $v_{p_i}^\text{ph}>v_{k_i}^\text{qb}$. 
    }
    \label{fig:v}
\end{figure}

Let us characterize the scattering for the specific case of cosine-like dispersion relations, Eqs. \eqref{eq:wp} and \eqref{eq:xik}. In this situation, as detailed in Appendix \ref{app:s_matrix}, the S-matrix can be written in the form 
\begin{equation}\label{eq:S_delta}
\begin{split}
    S_{f,i} &= \delta(p_i+k_i-p_f-k_f)
    \\
    &
    \times\bigg[t(k_i,p_i) \delta(p_f-p_i) + r(k_i,p_i) \delta(p_f - p_{f,2})\bigg].
\end{split}
\end{equation}
Here, the first Dirac delta establishes momentum conservation, whereas the two terms inside the brackets describe the two possible scattering processes (Fig. \ref{fig:v}a): on the one hand, an \emph{elastic} transmission, with probability amplitude $t(k_i,p_i)$, where both the scattered photon and the qubit conserve their momentum. On the other hand, an \emph{inelastic} process with probability amplitude $r(k_i,p_i)$, where the momenta of both photon and qubit change, $\{p_i,k_i\} \to \{p_{f,2},p_i+k_i-p_{f,2}\}$. These two possible scattering outcomes stem from the two solutions of the energy and momentum conservation equations,
\begin{align}
    \tilde{\omega}_{k_f+p_f,p_f} = & \tilde{\omega}_{k_i+p_i,p_i}, \label{eq:conservation_energy} \\
    k_f+p_f = & k_i+p_i, \label{eq:conservation_momentum}
\end{align}
which have a trivial solution $p_f=p_i$ and a second solution given by (see App. \ref{app:s_matrix} for details)
\begin{align}
 p_{f,2} &= \text{sign}\Big\lbrace\vert z(K_i)\vert^2\sin p_i+(v_{k_i}^{\text{qb}}-v_{p_i}^{\text{ph}})\text{Re}[z(K_i)]\Big\rbrace \nonumber \\
 &\times\arccos\left[\cos p_i -J'\sin (K_i)\frac{v_{k_i}^{\text{qb}}-v_{p_i}^{\text{ph}}}{\vert z(K_i)\vert^2}\right],\label{eq:pf2}
\end{align}
where the group velocities $v_p^{\text{ph}}$ and $v_k^{\text{qb}}$ are given by Eqs. \eqref{eq:vph} and \eqref{eq:vqb} respectively, and we have defined $K_i\equiv k_i+p_i$ and
\begin{align}
     & z(K) =  J+J' e^{-iK}.\label{eq:z_def}
    %\\
    % & Q(p,K) =  J\sin p + J'\sin(p-K).\label{eq:Q_def}
\end{align}
Note that, in the static-qubit limit $J'=0$, we recover $p_{f,2} = -p_i$, so that the second process becomes an elastic reflection, as reported in the literature \cite{Shen2005a,Shen2005b,Zhou2008a}.
Indeed, although the momentum $p_{f,2}$ can take both positive or negative values, the inelastic process can be understood as a reflection \emph{from the reference frame of the qubit}, as the corresponding group velocity of the scattered photons always lies below the group velocity of the scattered qubit. To illustrate this, we display in Fig. \ref{fig:v}b the group velocities of photons and qubits, namely Eqs. \eqref{eq:vph} and \eqref{eq:vqb}, evaluated at both allowed final momenta. This figure evidences that, in the usual scattering scenario $v_{p_i}^{\rm ph}>v_{k_i}^{\rm qb}$ the transmitted photon always moves faster than the qubit, whereas the reflected photon always remains behind the qubit, i.e. $v_{p_{f,2}}^{\rm ph} < v_{k_{f,2}}^{\rm qb}$. An equivalent argument holds for the case $p_i<0$, see Fig.~\ref{fig:v_appendix} in Appendix~\ref{AppendixAdditionalFigures} for details.

\subsection{Scattering amplitudes: nonreciprocity and motional energy reduction.}\label{SubsecAmplitudes}

Once the scattering processes  are characterized, we focus on their respective probability amplitudes.
The coefficients $t(k_i,p_i)$ and $r(k_i,p_i)$ read (see App. \ref{app:s_matrix})
\begin{align}
    &t(k_i,p_i) = \frac{\tilde{\omega}_{k_i+p_i,p_i}-(\Delta+\xi_{k_i+p_i})}{\tilde{\omega}_{k_i+p_i,p_i}-(\Delta+\xi_{k_i+p_i}) + i \frac{\Omega^2}{2J\sin{p_i}-2J'\sin{k_i}}}, \label{eq:t}\\
    &r(k_i,p_i) = \frac{-i \frac{\Omega^2}{2J\sin{p_i}-2J'\sin{k_i}}}{\tilde{\omega}_{k_i+p_i,p_i}-(\Delta+\xi_{k_i+p_i}) + i \frac{\Omega^2}{2J\sin{p_i}-2J'\sin{k_i}}}.\label{eq:r}
\end{align}
We see that, in agreement with
our interpretation of Eq. \eqref{eq:H1K}, these expressions are analogous to the case of a static qubit coupled to a linear cavity array with cosine dispersion relation \cite{Zhou2008a} under (i) replacement of the dispersion relation of the array by the total input energy $\tilde{\omega}_{k_i+p_i,p_i}$, (ii) addition of the shift $\xi_{k_i+p_i}$ to the level splitting $\Delta$, and (iii) substitution of the denominator of the third term of each denominator by the density of states of the effective dispersion relation $\tilde{\omega}_{k_i+p_i,p_i}$. The scattering coefficients fulfill $|t(k_i,p_i)|^2+|r(k_i,p_i)|^2=1$, which ensures that there are no more scattering channels \footnote{Note that the conservation of probability cannot be strictly obtained with the effective theory for high kinetic energies \cite{Calajo2017} since, by ignoring both the quantum fluctuations and the photon recoil of the qubit motion, the resulting light-qubit system becomes an open system.}. Because of this constrain, it is sufficient to focus on the transmission probability $|t(k_i,p_i)|^2$ in order to to characterize the scattering.

\begin{figure}
    \centering
    \includegraphics[width=\linewidth]{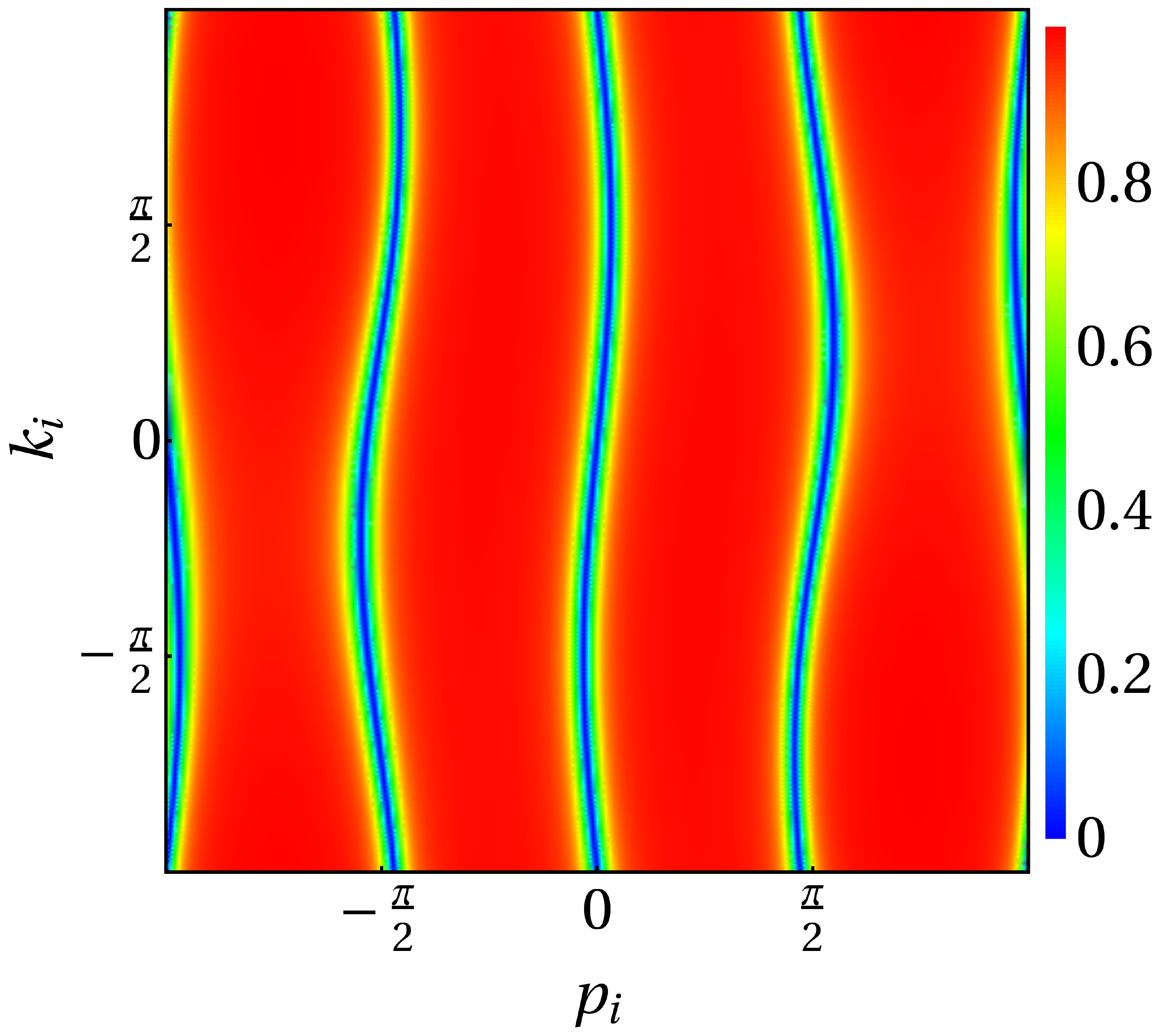}
    \caption{Transmission probability, $|t(k_i,p_i)|^2$, as a function of $p_i$ and $k_i$ for $\Omega=0.5J$, $J'=0.1J$, and $\Delta=0$.
    }
    \label{fig:t_lowJ}
\end{figure}

Before describing the effects of the moving qubit onto the scattering, we summarize here the properties of $|t(k_i,p_i)|^2$ in the static limit, $J'=0$ (see \cite{Zhou2008a}). In such a case, (i) transmission tends to 1 as the photon and the qubit are off resonance, (ii) there is full reflection under the resonant condition $\omega_{p_i}=\Delta$, (iii) there is no transmission at the band limits, $p_i=0$ and $\pm\pi$, since the group velocity of the photon is 0, so there is no propagation, and (iv) $|t(k_i,p_i)|^2$ is an even function of $p_i$, resulting in symmetric right-to-left and left-to-right scattering outcomes. For the moving qubit case, we show in Fig. \ref{fig:t_lowJ} the transmission probability $|t(k_i,p_i)|^2$ as a function of $k_i$ and $p_i$, for a moderated value of $J'$ and taking a value for the energy gap lying in the middle of the photonic band $\omega_p$ ($\Delta=0$). The deviations with respect to the static case are evident. In particular, full-reflection resonances $\vert t(k_i,p_i)\vert^2=0$, which in the static case appear at $p_i=\pm\pi/2$ are now shifted and depend on the initial qubit momentum $k_i$. This is due to the fact that the momenta for which full reflection occurs, namely the solutions of $\omega_{p_i}+\xi_{k_i} = \Delta + \xi_{k_i+p_i}$ (see numerator of Eq.~\eqref{eq:t}), are generally different from those in the limit $J'=0$. Also, the abscence of transmission when the velocity of the incident photon vanishes, which ocurrs at $p_i=\pm\pi$ in the static case, is shifted too; now this happens when both initial velocities, $v_{p_i}^\text{ph}$ and $v_{k_i}^\text{qb}$, are equal. Finally, $|t(k_i,p_i)|^2$ is not an even function of $p_i$ anymore, indicating a directional scattering. In other words, the scattering amplitudes for photons propagating with momenta $p_i$ and $-p_i$ are different, since the initial state of the system breaks time-reversal invariance for non-zero initial qubit momentum. 
Motion-induced reciprocity breaking stems from the Doppler effect, and is thus a common feature of systems with moving components such as trapped ions, where such asymmetries can be harnessed e.g. for motional cooling \cite{WiemanRMP1999}.
In the context of waveguide QED, it
has been demonstrated through directional emission both for qubits moving along classical trajectories \cite{Calajo2017}, and for static emitters in the presence of acoustic waves \cite{CalajoPRA2019} or moving conductors \cite{PratCampsPRL2018}. However, nonreciprocal single-photon scattering has so far been only reported in chiral waveguide systems with \emph{static} qubits \cite{LodahlNature2017}, where the non-reciprocity stems from the combined effect of photonic spin-orbit coupling and the intrinsic time-reversal symmetry breaking of the \emph{internal} qubit states (e.g. through spin-defined optical transitions). The quantum-motion-based nonreciprocal scattering demonstrated in this paper thus represents an alternative route to non-reciprocal single photon scattering, and might lead to the design or improvement of few-photon non-reciprocal devices such as diodes, transistors, or circulators \cite{RoyPRB2010,SayrinPRX2015,YanEPL2015,GonzalezBallesteroPRA2016}.

\begin{figure}
    \centering
    \includegraphics[width=\linewidth]{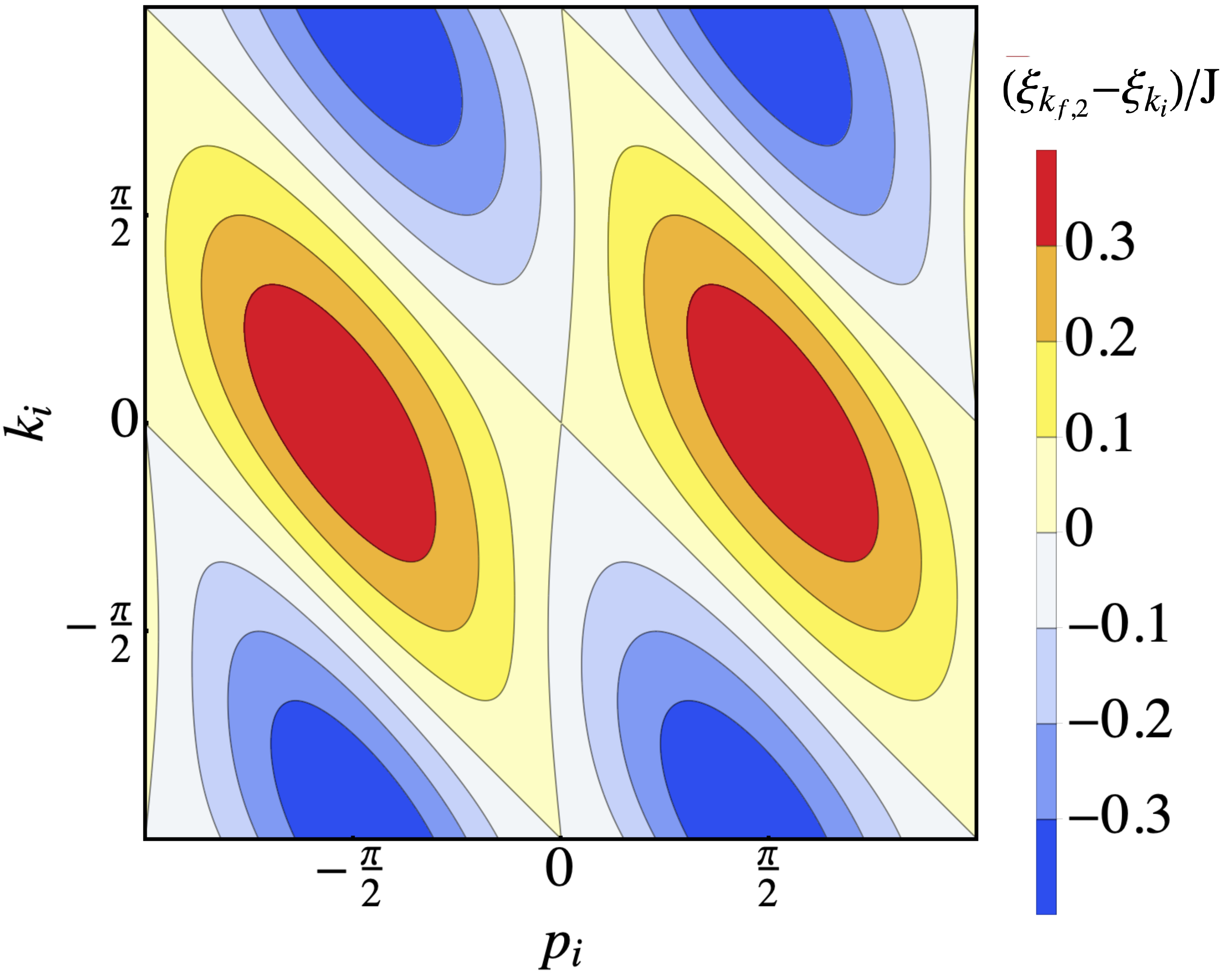}
    \caption{Difference between the final and the initial qubit energy after the reflection scattering process in units of $J$, namely $(\xi_{k_{f,2}}-\xi_{k_i})/J$, as a function of $p_i$ and $k_i$ and for $J'=0.1J$.  The blue and red regions correspond to reduction and increase of the qubit motional energy, respectively.}
    \label{fig:cooling}
\end{figure}

The above discussions on non-reciprocity concern the modification of the photonic state by influence of the qubit. However, we can as well exploit the rich scattering dynamics to control the motional state of the qubit via single-photon scattering. Specifically, in the reflection process, the final energies of both the photon and the qubit motion are modified as $p_{f,2}\neq -p_i$ (see Eq. \eqref{eq:pf2}). In Fig. \ref{fig:cooling}, we display the ($\Delta$-independent) energy variation experienced by the qubit during this scattering process, $\xi_{k_{f,2}}-\xi_{k_i}$, as a function of the photon and qubit initial momenta, $p_i$ and $k_i$ respectively. Remarkably, the motional energy of the qubit is reduced on a wide range of values of momenta $p_i,k_i$, indicating a single-photon-scattering-based energy depletion of the qubit motion. This depletion is maximized for initial photon momenta near the linear part of the band ($p_i\approx\pm\pi/2$) and high initial qubit energies ($k_i\approx \pm\pi$), a property that extends to other, and possibly to all, forms for the dispersion relations \footnote{{For instance one can check, by deriving the corresponding $S-$matrix, that  for a free qubit ($\xi_k\propto k^2$) in a waveguide with a linear dispersion relation ($\omega_p \propto \vert p \vert$), the energy depletion is also maximally efficient at large initial qubit momenta $k_i$.}}  
Moreover, one can choose $\Delta$, $J$, and $J'$ such that full reflection coincides with the area of maximum depletion (see Fig. \ref{fig:t_lowJ}), thus resulting in deterministic reduction  of the qubit motional energy. 
% One can demonstrate that such deterministic character can be attained for any initial qubit momentum $k_i$ by tuning both $\Delta$ and $p_i$.
Finally, note that for qubits with a given initial momentum distribution the energy depletion can remain efficient and practically deterministic, if the parameters are chosen to include all, or most, of such momentum distribution within the energy depletion region (i.e. the blue region) in Fig.~\ref{fig:cooling}.
 We emphasize that deterministic depletion of the qubit motional energy can be attained for any qubit and photon dispersion relations, and is thus applicable to any waveguide QED platform. Moreover, such motional depletion could in principle be optimized in each specific setup, i.e. for each particular qubit dispersion, e.g. by engineering the photonic band to minimize the final motional energy of the qubit.

 The mechanism for energy depletion observed in  Fig.~\ref{fig:cooling}, namely recoil of a scattered photon, is analogous to the radiation pressure mechanism in trapped ions, which has been exploited for dissipative laser cooling~\cite{WiemanRMP1999} or for recoil-induced lasing without population inversion~\cite{HoltPRA1977,GheriPRL1995}. In this case, the modification of the motional energy is attained through controlled scattering of guided photons. Hence, this process might be an interesting addition to the toolbox in experiments with qubits coupled to photonic structures, where additional control could be used to compensate unavoidable nanostructure-induced heating mechanisms~\cite{HummerPRX2019}.

\section{Bound states}\label{SecBoundStates}

Photonic baths with a band gap coupled to qubits support the existence of bound states \cite{Khalfin1958,Bykov1975,John1984,John1987,John1990,John1991,John1994}, where a photonic cloud is tightly localized around the impurities. 
These states have attracted significant interest in the community of wQED as they can be used to efficiently mediate qubit-qubit interactions \cite{Tudela2015,Shi2016,Calajo2016,SanchezBurilloArxiv2019,Bello2019}.
The single-particle bound states of the Hamiltonian have been already characterized in the literature when the qubit is \emph{static} ($\partial_k\xi_k = 0$) \cite{Garmon2009,Longo2010,Longo2011,Garmon2013,Lombardo2014,Shi2016,Calajo2016,Sanchez-Burillo2017}, even beyond the rotating-wave approximation \cite{Sanchez-Burillo2014}. In this section, we study these bound states for a moving qubit. First, we compute these states in momentum representation and characterize their existence conditions in Sec. \ref{SubsecKrepresentation}. Then, we analyze in Sec. \ref{SubsecPOSrepresentation} the representation of the bound states in position space and its difference with respect to the static qubit case. Finally,  we determine the bound-state energies and compute and discuss their dispersion relation as a function of total momentum $K$ in Sec. \ref{subsecExoticdispersions}.

\subsection{Bound states in momentum representation and existence conditions}\label{SubsecKrepresentation}

Let us firstly define the bound states and the eigenstates in the continuum. Generally, a single-excitation eigenstate of the Hamiltonian $H_{1,K}$ in Eq.~\eqref{eq:H1K} with total momentum $K$ can be written as
\begin{equation}\label{eq:psik}
    \ket{\Psi_K} = u_K \ket{K}  + \sum_p f_K(p) \ket{p}_K,
\end{equation}
where we are using the states introduced in Eq.~\eqref{eq:states}.
If the energy of $\ket{\Psi_K}$ is $\mathcal{E}_K$, we can obtain the coefficients in Eq. \eqref{eq:psik} by solving the time-independent Schr\"odinger equation to obtain 
\begin{align}
    \label{eq:fkp}f_K(p) = & \frac{\Omega}{\sqrt{L}}\frac{u_K}{\mathcal{E}_K-\tilde{\omega}_{K,p}},\\
    \label{eq:ukm2} u_K^{-2} = & 1+\frac{\Omega^2}{L}\sum_p \frac{1}{(\mathcal{E}_K-\tilde{\omega}_{K,p})^2}.
\end{align}
Additionally, one can obtain the following equation for the eigenvalue $\mathcal{E}_K$:
\begin{equation}
    \mathcal{E}_K =  E_{K,\Delta}+\frac{\Omega^2}{L}\sum_p \frac{1}{\mathcal{E}_K-\tilde{\omega}_{K,p}} \label{eq:Ek}.
\end{equation}
We emphasize that, in the above three expressions, we assume a fixed value of the constant of motion $K$, whereas $p$ is a free variable. In agreement with our interpretation of 
Eq. \eqref{eq:H1K}, the above expressions are analogous to those obtained for a non-moving qubit coupled to a bath \cite{Shi2016}, under substitution of the qubit energy by $E_{K,\Delta}$ (Eq.~\eqref{eq:impurityK}) and of the bath dispersion relation by $\tilde{\omega}_{K,p}$ (Eq. \eqref{eq:dispersionK}).
Equations \eqref{eq:fkp}-\eqref{eq:Ek} are valid for any single-excitation eigenstate of waveguide QED Hamiltonians of the type Eq.~\eqref{eq:H1K}. Hereafter we particularize our study to bound eigenstates of our system, namely eigenstates of the above form for which $\mathcal{E}_K \notin \tilde{\omega}_{K,p}$ with the dispersion relations Eqs.~\eqref{eq:wp} and \eqref{eq:xik}.

We now determine the existence conditions for bound states of $H_{1,K}$. We commence by considering the continuum limit $L\to \infty$ and defining the following function,
\begin{equation}\label{eq:F}
    F(E) \equiv E-E_{K,\Delta} - \Sigma_K(E),
\end{equation}
with $\Sigma_K(E)$ the self-energy
\begin{equation}\label{eq:self-energy}
    \Sigma_K(E)\equiv \frac{\Omega^2}{2\pi}\int_{-\pi}^\pi \frac{dp}{E-\tilde{\omega}_{K,p}}.
\end{equation}
The function $F(E)$ is defined for all $E \in \mathbb{R}$ such that $E\leq \text{min}(\tilde{\omega}_{K,p})$ and $E\geq \text{max}(\tilde{\omega}_{K,p})$, where $F(E)$ takes real values.
Bound states are solutions of the equation $F(E)=0$ with $E\notin \tilde{\omega}_{K,p}$ (cf. Eq. \eqref{eq:Ek}). Notice that $\lim_{E\to\pm\infty}F(E)=\pm\infty$. Besides, we can easily see that $dF(E)/dE>0$, so $F(E)$ is a monotonously increasing function. Then, a bound state $\ket{\Psi_{K,-}^\text{bs}}$ exists with energy $\mathcal{E}_{K,-}<\tilde{\omega}_{K,p}$ if and only if
\begin{equation}
    \lim_{E\to\text{min}\tilde{\omega}_{K,p}^-} F(E) >0
\end{equation}
 and a bound state $\ket{\Psi_{K,+}^\text{bs}}$ exists with energy $\mathcal{E}_{K,+}>\tilde{\omega}_{K,p}$ if and only if
\begin{equation}
    \lim_{E\to\text{max}\tilde{\omega}_{K,p}^+} F(E) <0.
\end{equation}
We compute the values of the above limits to obtain
\begin{equation}
    \begin{split}
        &\lim_{E\to\text{min}\tilde{\omega}_{K,p}^-} F(E) = +\infty, \\
        &\lim_{E\to\text{max}\tilde{\omega}_{K,p}^+} F(E) = -\infty,
    \end{split}
\end{equation}
independently of the values of the parameters. From this, we conclude that both bound states, $\ket{\Psi_{K,-}^\text{bs}}$ and $\ket{\Psi_{K,+}^\text{bs}}$, exist for all the values of the parameters $\Delta,\Omega,J,J',K$.

\subsection{Asymmetric phase in position space}\label{SubsecPOSrepresentation}

The calculation of the bound states allows us to compare them with their counterpart for a static qubit, specifically the potential differences in the wavefunction and in the bound-state energies. We will start by the former, where for the sake of clarity we calculate the wavefunction of the state Eq. \eqref{eq:psik},  for $\ket{\Psi_K}=\ket{\Psi_{K,\pm}^\text{bs}}$ in the position representation. By using the original notation in terms of creation and annihilation operators
(see e.g. Eq. \ref{eq:Hb}) in the second term in Eq. \eqref{eq:psik}, we cast the bound state wavefunction in the following form,
\begin{equation}
\begin{split}
    \ket{\Psi_{K,\pm}^\text{bs}} & = u_{K,\pm}^\text{bs} e_K^\dagger\ket{0} \\
    &+ \frac{1}{L}\sum_{p,x_1,x_2} e^{ip(x_1-x_2)} f_{K,\pm}^\text{bs}(p) e^{iKx_2} a_{x_1}^\dagger g_{x_2}^\dagger\ket{0},
\end{split}
\end{equation}
where $a^\dagger_x$ and $g^\dagger_x$, namely the Fourier transforms of the operators $a^\dagger_p$ and $g^\dagger_k$ respectively, are defined in the usual way \cite{ShenPRA2009}. These new operators correspond to creation of excitations, either photons or motional states, at a well defined position $x$.
Therefore, we can identify the Fourier transform of $f_{K,\pm}^\text{bs}(p)$~(Eq. \eqref{eq:fkp}), namely $\tilde{f}_{K,\pm}^\text{bs}(x) \equiv L^{-1/2}\sum_p e^{ipx} f_{K,\pm}^\text{bs}(p)$,
%, the Fourier transform of $f_{K,\pm}^\text{bs}(p)$ (Eq. \eqref{eq:fkp}), is proportional 
as the probability amplitude of detecting the ground-state qubit and the photon at a relative position $x\equiv x_{\rm photon}-x_{\rm qubit}$, aside from a proportionality constant, i.e., 
 $\tilde{f}_{K,\pm}^\text{bs}(x) \propto \bra{0}a_{x}g_{0}\ket{\Psi_{K,\pm}^\text{bs}}$. This probability amplitude reads (see App. \ref{app:bs} for details)
\begin{align}
\label{eq:fkxp}     \tilde{f}_{K,\pm}^\text{bs}(x)=
          \frac{\Omega u_{K,\pm}^\text{bs}}{J+J'e^{-iK}} \frac{y_<^x}{y_<-y_>} \quad & \text{if } x>0, \\
\label{eq:fkxm}     \tilde{f}_{K,\pm}^\text{bs}(x)=
          \frac{\Omega u_{K,\pm}^\text{bs}}{J+J'e^{iK}} \frac{(y_<^*)^{|x|}}{y_<^*-y_>^*}\quad & \text{if } x<0,
\end{align}
where $y_<$ and $y_>$ are respectively the poles of the self-energy inside and outside the unit circle, evaluated at $E=\mathcal{E}_K$ (see App. \ref{app:self-energy}). 
By using the explicit form of these functions and the properties of the bound states, we can cast this function in the compact form
\begin{equation}
    \tilde{f}_{K,\pm}^\text{bs}(x)=\frac{\Omega u_{K,\pm}^{\rm bs}}{\vert\sqrt{\mathcal{E}_K^2-4\vert z(K)\vert^2}\vert}\vert y_<\vert^{\vert x\vert}e^{-i\vert x\vert \text{Arg} [z(K)]},
\end{equation}
where $z(K)$ is defined by Eq. \eqref{eq:z_def}. From the above equation we deduce, first, that, the photonic cloud is exponentially localized around the position of the qubit, with a localization length equal to $-1/\text{Re}(\log(y_<))$. This localization is usual in such bound states and also appears in the static case. Moreover, and also like in the static qubit situation, the total probability density is symmetric around $x=0$, i.e., $|\tilde{f}_{K,\pm}^\text{bs}(x)|=|\tilde{f}_{K,\pm}^\text{bs}(-x)|$. However, the probability amplitudes are not, as the \emph{phase} changes sign with $x$. Such a phase depends on $J'/J$ and $K$ and vanishes both for the static qubit case $J'=0$ and within the subspaces $K=0,\pm\pi$. Although not relevant for a single qubit, this asymmetric phase has an impact on the mediated qubit-qubit interactions, and can be exploited to engineer complex interactions as recently reported for static qubits in complex nonreciprocal lattices \cite{SanchezBurilloArxiv2019}.
As a final remark, note that, since the qubit is fully delocalized for an eigenstate $\ket{\Psi_{K,\pm}^\text{bs}}$, also the photon is. Specifically, the photon number at position $x_0$, given by
\begin{align}
     &\bra{\Psi_{K,\pm}^\text{bs}} a_{x_0}^\dagger a_{x_0}  \ket{\Psi_{K,\pm}^\text{bs}} \\
     &= 
    \frac{|\Omega u_{K,\pm}^\text{bs} y_<|^2}{|(J+J'e^{iK})(y_<-y_>)|^2(1-|y_<|^2)}, \nonumber
\end{align}
\emph{does not} depend on the position $x_0$. This does not contradict the usual definition of bound state since, as explained above, the photon and the qubit remain exponentially bound to each other.

\subsection{Energies of the bound states: Exotic dispersion relations}\label{subsecExoticdispersions}

\begin{figure}[tbh!]
\includegraphics[width=\linewidth]{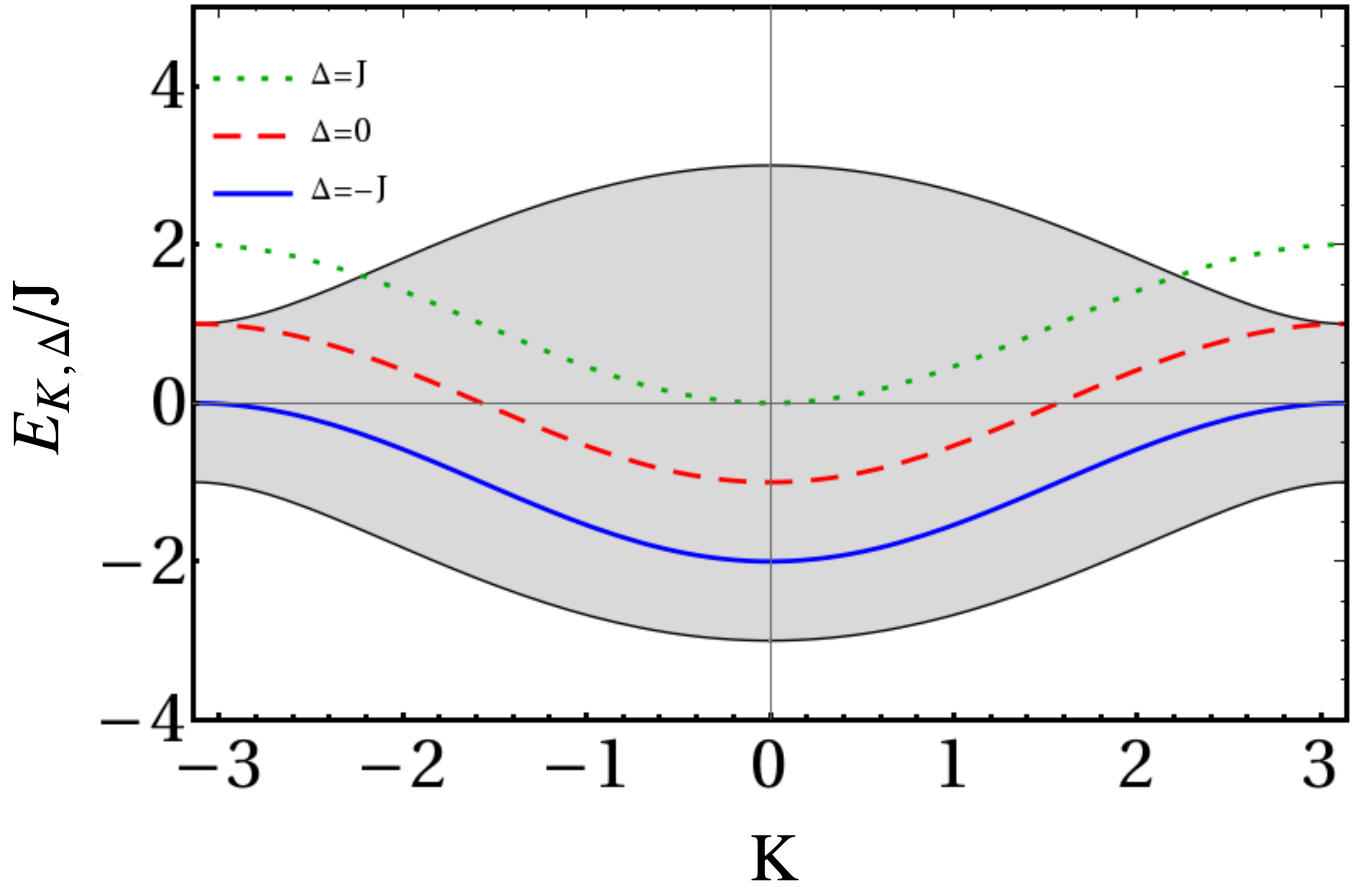}
\caption{Effective qubit energy in the subspace of momentum $K$,  $E_{K,\Delta}$, as a function of $K$, for $J'=0.5J$. The shaded area indicates the range of values taken by the effective band energy $\tilde{\omega}_{K,p}$.}
\label{fig:xik}
\end{figure}

Let us finally analyze the energies of the bound states.
These bound states are the result of a hybridization between the states $e_K^\dagger\ket{0}$ and $a_p^\dagger g_{K-p}^\dagger\ket{0}$, with energies $E_{K,\Delta}$ and $\tilde{\omega}_{K,p}$ respectively. In Fig. \ref{fig:xik} we display the energy of the first component of the hybrid bound state, namely $E_{K,\Delta}$, for different values of $\Delta$ and for $J'=0.5J$. In the shaded region of the same plot, we display the possible values of $\tilde{\omega}_{K,p}$, namely the energy of the second component of the bound state. As opposed to the static case, both these energies form a continuum as a function of the total momentum $K$ and, specifically, the excited-state contribution $E_{K,\Delta}$ is not a horizontal line.
As a consequence, the gaps between $E_{K,\Delta}$ and the upper and lower limits of the band $\tilde{\omega}_{K,p}$ are different for each value of $K$, even at $\Delta=0$. Therefore, as opposed to the case of a static qubit \cite{Longo2011}, the energies of the two bound states will not be symmetric around the bands when the qubit energy lies in the middle of the band, $\Delta=0$.

\begin{figure}
    \centering
    \includegraphics[scale=0.3]{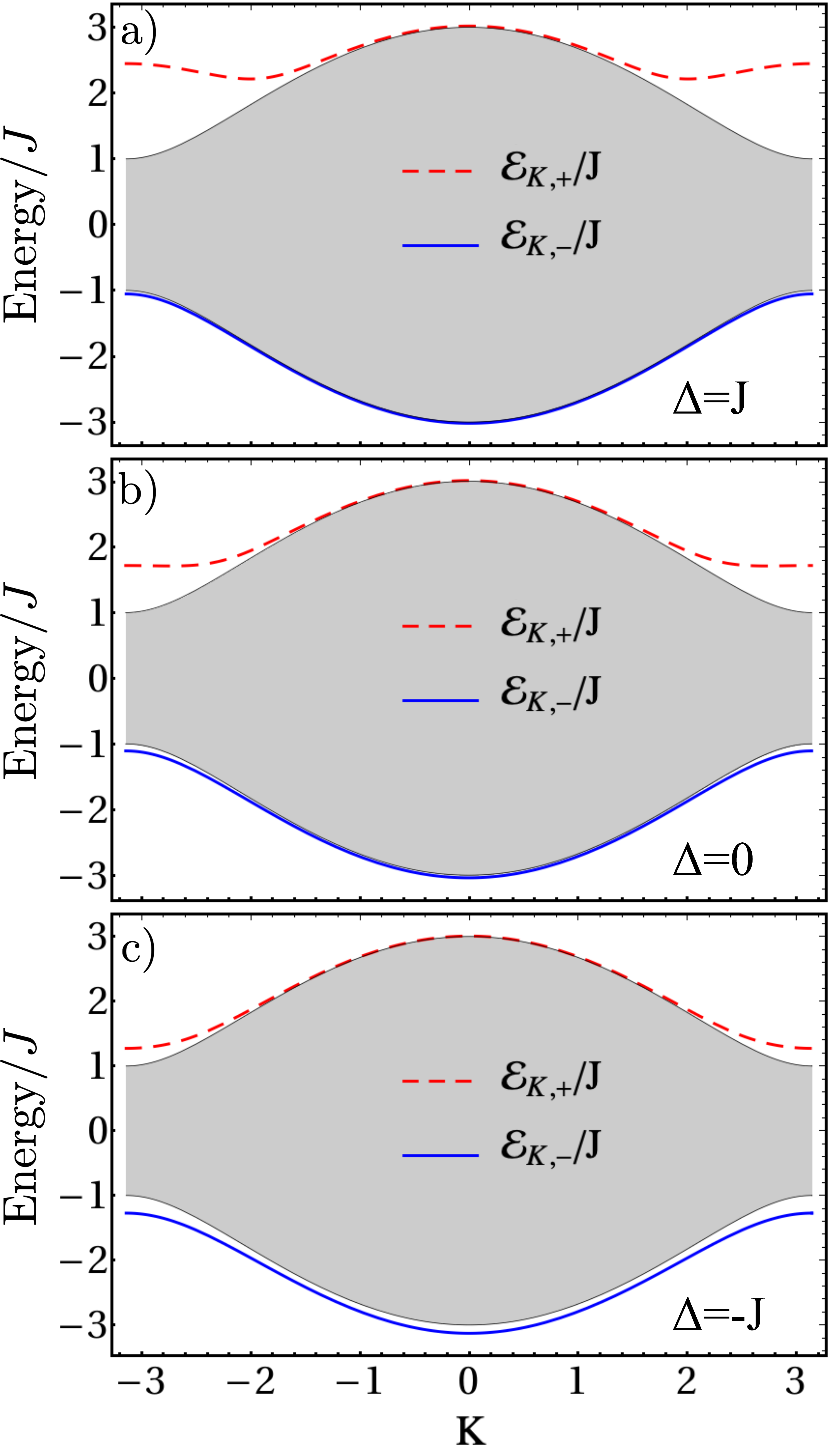}
    \caption{Energies of the bound states as a function of $K$. We choose $J'=0.5J$ and $\Omega=J$. The values of $\Delta$ are in the figures. The shaded region represents the values taken by $\tilde{\omega}_{K,p}$.}
    \label{fig:bs_energies}
\end{figure}

The energies of the two bound states, $\mathcal{E}_{K,\pm }$, are shown in Fig.~\ref{fig:bs_energies} as a function of total momentum $K$, for three different values of $\Delta$. The above predicted asymmetry at $\Delta=0$ is manifest in panel~\ref{fig:bs_energies}b). For a clearer picture, it is convenient to project the full Hamiltonian \eqref{eq:H} in the single-excitation subspace (but not for a fixed value of $K$) and span it in the bound $\ket{\Psi_{K,\pm}^\text{bs}}$ and scattering states $\ket{\Psi_{K,p}^\text{sc}}$ as
\begin{equation}
    H_1 = \sum_{p,K}\tilde{\omega}_{K,p} \ket{\Psi_{K,p}^\text{sc}} \bra{\Psi_{K,p}^\text{sc}} + \sum_{s=\pm 1,K} \mathcal{E}_{K,s}\ket{\Psi_{K,s}^\text{bs}} \bra{\Psi_{K,s}^\text{bs}}.
\end{equation}
From the above picture, we can think of the bound states as two bands with dispersion relation $\mathcal{E}_{K,\pm }$. As evidenced e.g. by the blue curves $\mathcal{E}_{K,- }$ in Fig. \ref{fig:bs_energies}a-c), these dispersion relations can inherit the typical cosine-like shapes from the photonic and motional bands. On the other hand, exotic band shapes can arise as well, which can give rise to interesting quantum dynamics. For instance, the dispersion relation of $\mathcal{E}_{K,+}$ in Fig. \ref{fig:bs_energies}a) displays two unusual minima at non-zero momentum $K$. These local minima in the dispersion relation have risen interest in disciplines such as cavity polaritonics and magnonics, for instance in the context of Bose-Einstein condensation  \cite{LiSciRep2013,LaiNature2007,CerdaPRL2010,TaneseNatComm2013}. On the other hand, at $\Delta=0$, panel \ref{fig:bs_energies}b), the band $\mathcal{E}_{K,+}$ tends to be flattened around the edges of the Brillouin zone, an indication of the band becoming non-linear as the quadratic contribution is cancelled, i.e., $\mathcal{E}_{K,+}\big\vert_{K=\pi} \sim (K-\pi)^4$. Such quartic minima are well-known in the literature, as
particle dynamics generated by non-quadratic Hamiltonians  display strongly non-Gaussian dynamics~\cite{WeissPRR2019}. Our results thus evidence that the qubit motion could represent an additional asset for band engineering in wQED systems.

\section{Spontaneous emission}\label{sect:se}

The spontaneous emission of a static qubit coupled to a waveguide has been thoroughly studied in the literature, finding, for instance, power-law and fractional decays \cite{JohnPRA1994,LonghiPRL2006,MartorellChapter2009,Garmon2013,Lombardo2014,Sanchez-Burillo2017}. It was also recently studied for a qubit moving along a classical trajectory at constant velocity~\cite{Calajo2017}. In this section we extend the study to a qubit whose motion is described quantum mechanically. First, in Sec.~\ref{subsecREGIMES}, we study the regimes of spontaneous emission for an excited-state qubit with well-defined momentum $K$. Then, in Sec.~\ref{subsecKSelective}, we extend these results to a qubit with an initial momentum distribution, and demonstrate a new regime of emission where only certain components $K$ decay into photons. Then, in Sec. \ref{subsecSpatial}, we analyze the spontaneous emission in real space, and compare the cases of a moving qubit to a static one.

\subsection{Regimes of spontaneous emission for well-defined initial momentum $K$}\label{subsecREGIMES}

Our general aim is to determine the time evolution of an initially excited qubit with a given momentum distribution, i.e., of an initial state
\begin{equation}\label{eq:generalPsi0}
    \vert \Psi(t=0)\rangle = \sum_K\varphi_Ke^\dagger_K\vert 0 \rangle \equiv \sum_K\varphi_K\vert K \rangle.
\end{equation}
Two cases are of particular interest for this work: first, a qubit with a well-defined momentum $\varphi_K\propto \delta_{KQ}$, whose simpler solution is the building block for determining the dynamics of a general state Eq. \eqref{eq:generalPsi0}. Second, the initial state corresponding to the usual spontaneous emission scenario, namely a qubit initially localized at a given position $x_0$, i.e., $\varphi_K = e^{iKx_0}/L^{1/2}$ (note that hereafter any position $x$ is assumed to be given in units of the lattice spacing). 
Since in this case the analytical study of the dynamics is not possible, when considering such state we have solved its dynamical evolution by exact numerical diagonalization of the Hamiltonian.
In this section, we focus on the general aspects of the spontaneous emission for an initial state with well-defined momentum, i.e., $\vert\Psi(t=0)\rangle = \vert K\rangle$. Since both excitation number and linear momentum are conserved, the general expression of this state at time $t$ is given by
\begin{equation}\label{eq:psikEmission}
    \ket{\Psi_K}(t) = \psi_{eK}(t) \ket{K}  + \sum_p \phi_K(p,t) \ket{p}_K.
\end{equation}
The first contribution, $\psi_{eK}(t)$, is the excited-state probability amplitude of the qubit, which in the subspace of momentum $K$ has an effective energy $E_{K,\Delta}=\Delta+\xi_K$ (see Eq. \ref{eq:H1K}). Conversely, the second contribution $\phi_K(p,t)$ corresponds to the photonic part of the state, which propagates in a modified dispersion $\tilde{\omega}_{K,p}=\omega_p+\xi_{K-p}$.

\begin{figure}[tbh!]
\includegraphics[width=\linewidth]{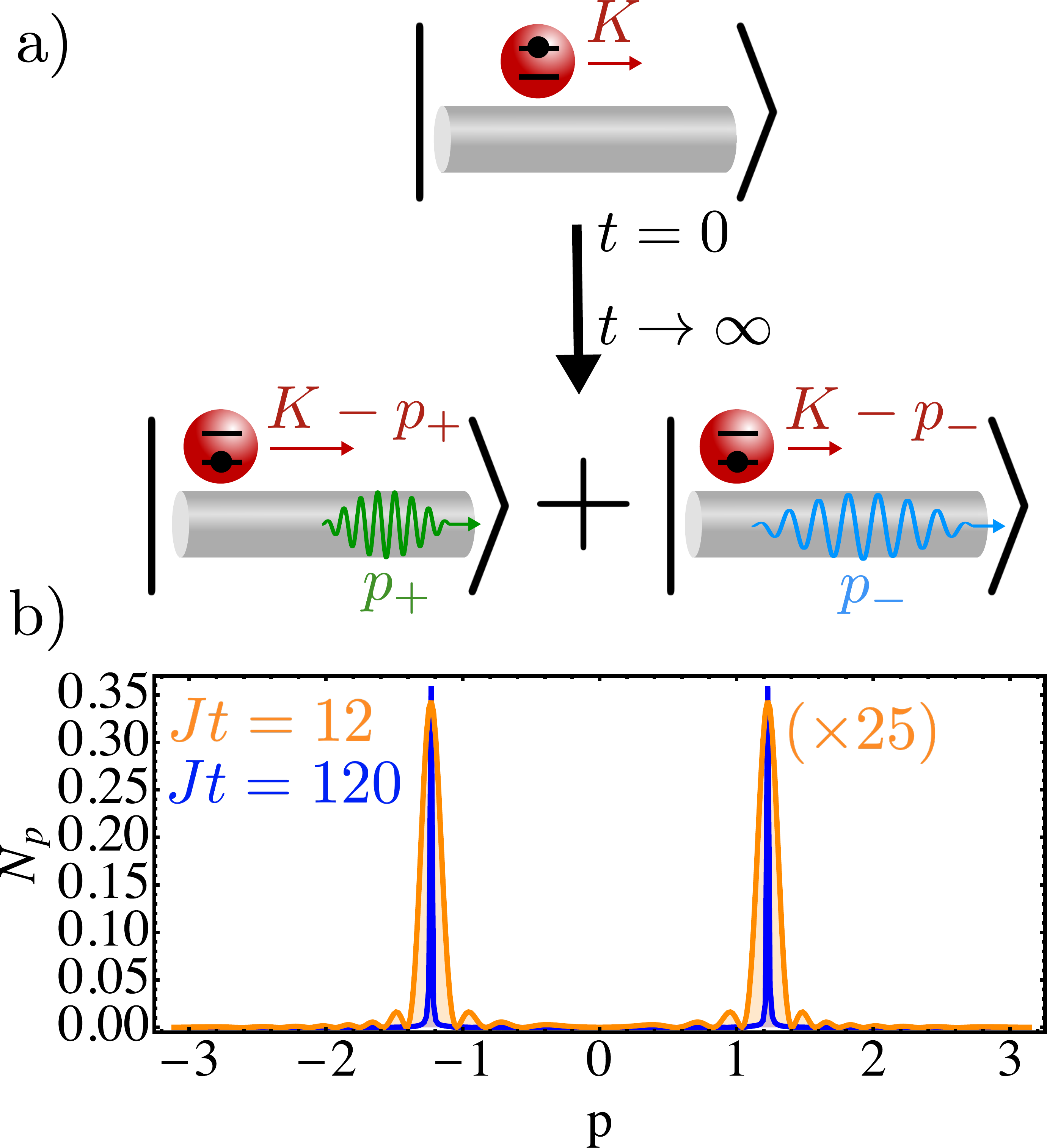}
\caption{a) Schematic representation of the spontaneous emission process for qubits embedded in the photonic band (see main text for details): an initially excited qubit with momentum $K$ decays into an entangled photon-motional state. b) Occupation of photon mode $p$ at short and long times (orange and blue lines, respectively), for $K=0$, $J'=0.5J$, $\Omega=0.2J$, and $\Delta=0$. The solutions of Eq. \eqref{eq:ppm} for these parameters are $p_\pm = \pm \arctan 2\sqrt{2} \approx \pm 1.23$.}
\label{fig:emission_K}
\end{figure}

Although we compute the dynamics of the above state in an exact manner through numerical calculations, it is convenient, in order to gain insight into the relevant physics, to characterize theoretically the emission in the Born-Markov regime; that is, we consider that the photonic degrees of freedom evolve much faster than those of the qubit, which allows us to adiabatically eliminate the photons~\cite{Cohen-Tannoudji1992,Gardiner2000}. Within this approximation, the qubit spontaneously decays at a rate equal to the density of states of the effective band $\tilde{\omega}_{K,p}$ at the effective qubit energy  $E_{K,\Delta}$. This allows us to identify the two well-known emission regimes: (i) If the energy of the qubit lies outside the band, i.e., if the equation $E_{K,\Delta} = \tilde{\omega}_{K,p}$ is not fulfilled for any $p$, then the qubit will not decay, and $\lim_{t\to\infty} \vert\psi_{eK}(t)\vert\to 1$. (ii) Conversely, if the qubit energy is embedded in the band, i.e., if $E_{K,\Delta} = \tilde{\omega}_{K,p}$ for some photonic momentum $p$, then the qubit will decay exponentially into guided photons, and $\lim_{t\to\infty} \vert\psi_{eK}(t)\vert\to 0$. In the latter regime, the momenta of the emitted photons at infinite time can be determined by energy conservation,
\begin{equation}\label{eq:energy_conservation}
    E_{K,\Delta} = \tilde{\omega}_{K,p}.
\end{equation}
This equation has two solutions, given by
\begin{align}\label{eq:ppm}
    & p_\pm = \arctan \left(\frac{-2\text{Im}[z^2(K)] 
    \pm E_{K,\Delta}\sqrt{4\vert z^2(K)\vert-E_{K,\Delta}^2}}{E_{K,\Delta}^2-4J'^2 \sin^2 K}\right),
\end{align}
where the branch of the function $\arctan(z)$ must be chosen to ensure that $p_\pm$ fulfills Eq. \eqref{eq:energy_conservation}. Note that in general $\text{sign}[p_+]=-\text{sign}[p_-]$, so that the emitted photonic wavepacket propagates in both right and left directions \footnote{One should not interpret this statement too rigorously, as the concept of direction is ill defined on a periodic lattice. Indeed, any positive momentum is equivalent to a negative one after substraction of an integer multiple of $2\pi$.}. The system state at long times can thus be written, in this regime, as a superposition of two product states,
\begin{equation}
\begin{split}
    \vert\Psi(t\to\infty)\rangle 
    =
    \sum_{p=p_+,p_-}\phi_{K\infty}(p)\vert p\rangle_K ,
\end{split}
\end{equation}
as schematically depicted in Fig. \ref{fig:emission_K}a). This is indicated also by the momentum distribution of the emitted photons, $N_p=\langle a^\dagger_pa_p\rangle$ (Fig. \ref{fig:emission_K}b)), which becomes sharply peaked around $p=p_\pm$ as time increases. Note that at short times (red curve) the distribution is slightly asymmetric with respect to the main peak at $p_\pm \approx \pm1.23$ as,
for $\vert p_{\pm} \vert < \pi/2$, the photonic density of states is higher below $p_{\pm}$ than above it, thus slightly favouring emission at low photonic momenta (see the discussion of Fig.~\ref{fig:emission_K_appendix} in Appendix~\ref{AppendixAdditionalFigures} for more details).

Since the photonic wavepacket emitted by a qubit with initial momentum $K$ contains both positive and negative momenta $p_\pm$, one might expect that the emitted photons would acquire chirality or directionality, that is, that there would be an imbalance between the intensity emitted at $p=p_+$ and the intensity emitted at $p=p_-$. This imbalance is suggested by the time-reversal asymmetry of the initial state $\vert K \rangle$ or, in a semiclassical picture, from the Doppler-shifted photonic density of states perceived from the perspective of a moving qubit. Indeed, we note that the state at infinite time, schematically depicted in Fig. \ref{fig:emission_K}a), is an entangled state, each of which components is strongly directional, as it describes a photon propagating with a single, well defined momentum plus the corresponding recoil-decelerated qubit in its ground state.  Moreover, since in general $|p_+|\neq |p_-|$ (except for $K=0$ or $K=\pm\pi$), the group velocities for $p_+$ and $p_-$ (Eq. \eqref{eq:vph}) are different in modulus, which suggests a directional emission. However, by quantifying the directionality of the emission in the usual way, namely through a directionality parameter
\begin{equation}
\begin{split}
    D &\propto \lim_{t\to\infty}\Big[\sum_{p>0} N_p - \sum_{p<0} N_p\Big] =
    \\
    &=\lim_{t\to\infty}[N_{p_+}-N_{p_-}] = \sum_{p=p_+,p_-}\text{sign}[p]\vert\phi_{K\infty}(p)\vert^2,
\end{split}
\end{equation}
one finds that, counterintuitively, the emission is not directional, i.e. $D=0$.
Indeed, the probabilities of emitting a photon with momentum $p_+$ and with momentum $p_-$, namely $\vert \phi_{K\infty}(p_+)\vert^2$ and $\vert \phi_{K\infty}(p_-)\vert^2$,
are equal, as the decay rate at momentum $p$, $\propto 1/(\partial\tilde{\omega}_{K,p}/\partial p)^2$, is the same for both $p_+$ and $p_-$ (see Fig.~\ref{fig:emission_K_appendix} in Appendix~\ref{AppendixAdditionalFigures} for more details). This stems from the fact that the left-hand side of Eq. \eqref{eq:energy_conservation} is $p$-independent, so that both $\tilde{\omega}_{K,p}$ and all its derivatives with respect to $p$ are identical for $p=p_+$ and $p=p_-$.
Note that the absence of directionality is a particular feature of the quantum nature of the qubit motion. Indeed, in the presence of decoherence or any other agent inducing the collapse of the motional wavefunction, such as e.g. external forces that keep the velocity of the qubit constant in time \cite{Calajo2017}, the emitted photons could acquire a  directional character.

As a final remark, let us briefly discuss the consequences of going beyond the Markovian approximation. Based on the fact that the self-energy is qualitatively equal to that of the static case (see App.~\ref{app:self-energy}), the non-Markovian effects shall be the same. After undergoing an initial exponential decay, the population of the qubit in the excited state with momentum $K$ will have a power-law decay \cite{Garmon2013,Sanchez-Burillo2017}. Finally, it will converge to the contribution of the bound states, which depicts oscillations with frequency given by the energy gap between both bound states, $\mathcal{E}_{K,+}-\mathcal{E}_{K,-}$ \cite{Lombardo2014,Sanchez-Burillo2017}, and a constant given by the overlap between the initial state and the bound states.

\subsection{K-selective emission}\label{subsecKSelective}

Let us now consider the spontaneous emission from an initial wavepacket of the form Eq.~\eqref{eq:generalPsi0}, i.e. with contributions from different momenta $K$. Since each of the components $\vert K\rangle$ evolves trivially, the two emission regimes described above also arise for such a state, namely a complete suppression of the emission if the qubit effective energies $E_{K,\Delta}$ lie outside of the effective band $\tilde{\omega}_{K,p}$ for all the values of $K$ fulfilling $\varphi_K \ne 0$, and a complete decay into guided photons if such energies are embedded in the band. Furthermore, since the effective energy depends on the wavevector $K$,
a third regime can arise in this case, where some of the components $\vert K \rangle$ will decay while others will not. This phenomenon of $K-$selective emission will occur when $\Delta$ is close to the maximum or the minimum of the photonic band $\pm2J$.

\begin{figure}[tbh!]
\includegraphics[width=\linewidth]{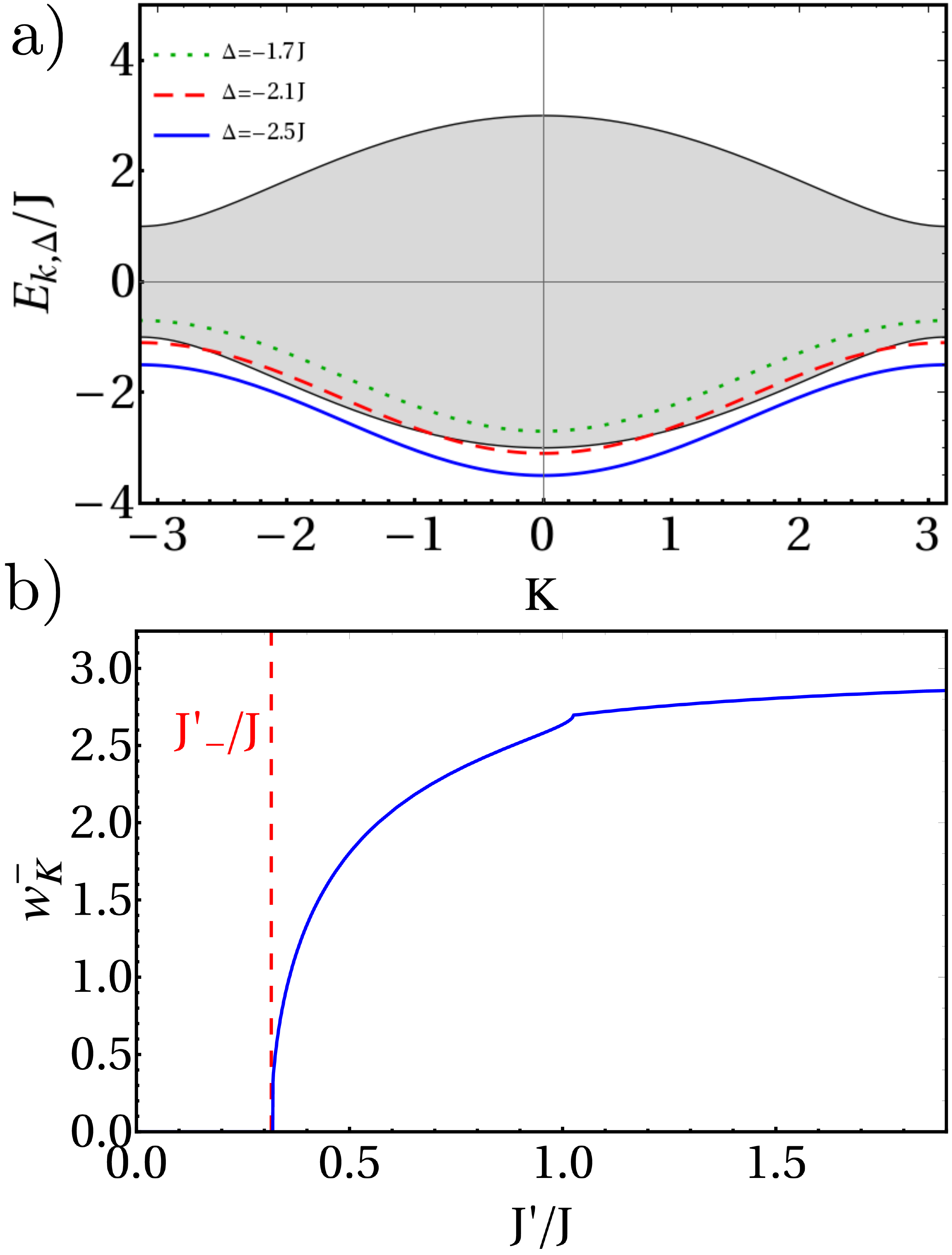}
\caption{a) Energy of the effective qubit $E_{K,\Delta}$, for $J'=0.5J$ and $\Delta=-1.7J,-2.1J,-2.5J$ (dotted green, dashed red, and solid blue, respectively), as a function of $K$. The shaded region renders the effective band $\tilde{\omega}_{K,p}$. b) Width of the window of momenta for which the qubit undergoes spontaneous emission (see main text for details) as a function of $J/J'$, for $\Delta=-2.1J$. The red dashed line indicates the critical value $J_-'\approx \sqrt{0.1}J$.
Notice that $w_K^-$ is defined here as the width of the window for positive momentum $K$; there is an equivalent window for negative values of $K$.}
\label{fig:xi_low}
\end{figure}

Even in the Markov regime, the analysis of the above described $K-$selective emission is not straightforward. We will focus here on the results, and address the reader to the approximate quantitative description given in Appendix \ref{app:perturbative}. Let us first focus on the case where the qubit frequency $\Delta$ is close to the minimum of the static band, $-2J$.  In Fig. \ref{fig:xi_low} we show the effective qubit energy $E_{K,\Delta}$ as a function of $K$ for three different values of $\Delta$ and $J'=0.5 J$, superimposed to the range of energies of the effective photonic band $\tilde{\omega}_{K,p}$ (gray area). Usual spontaneous emission regimes, namely full suppression of the decay and full decay into guided photons, 
correspond to the blue and green-dotted curves, respectively. 
%On the one hand, when $\Delta$ is far below $\omega_p$ (blue line) the resonance condition will never be satisfied and the qubit will not decay. On the other hand, when $\Delta$ is fully embedded in the band (green dotted line), the resonance condition is met for all $K$ and the qubit will spontaneously decay to its ground state by photon emission. 
However, there is an intermediate regime for which $E_{K,\Delta}$ is embedded in the band only for some values of $K$. In such regime, whether the qubit decays or not depends on its initial momentum. As detailed in Appendix \ref{app:perturbative}, this regime is characterized by $\Delta<-2J$ and $J'$ larger than a critical value $J_-'$, which in the limit $-\Delta-2J\ll J$ can be approximated as $J_-' \approx \sqrt{-J(\Delta+2J))}$. In this limit, the qubit will emit photons only for values of $K$ within a given interval $w_K^-$ centered at $K=\pm \pi/2$. As shown in Fig. \ref{fig:xi_low}b), the width of this emission window depends on the ratio $J'/J$.

\begin{figure}[tbh!]
\includegraphics[width=\linewidth]{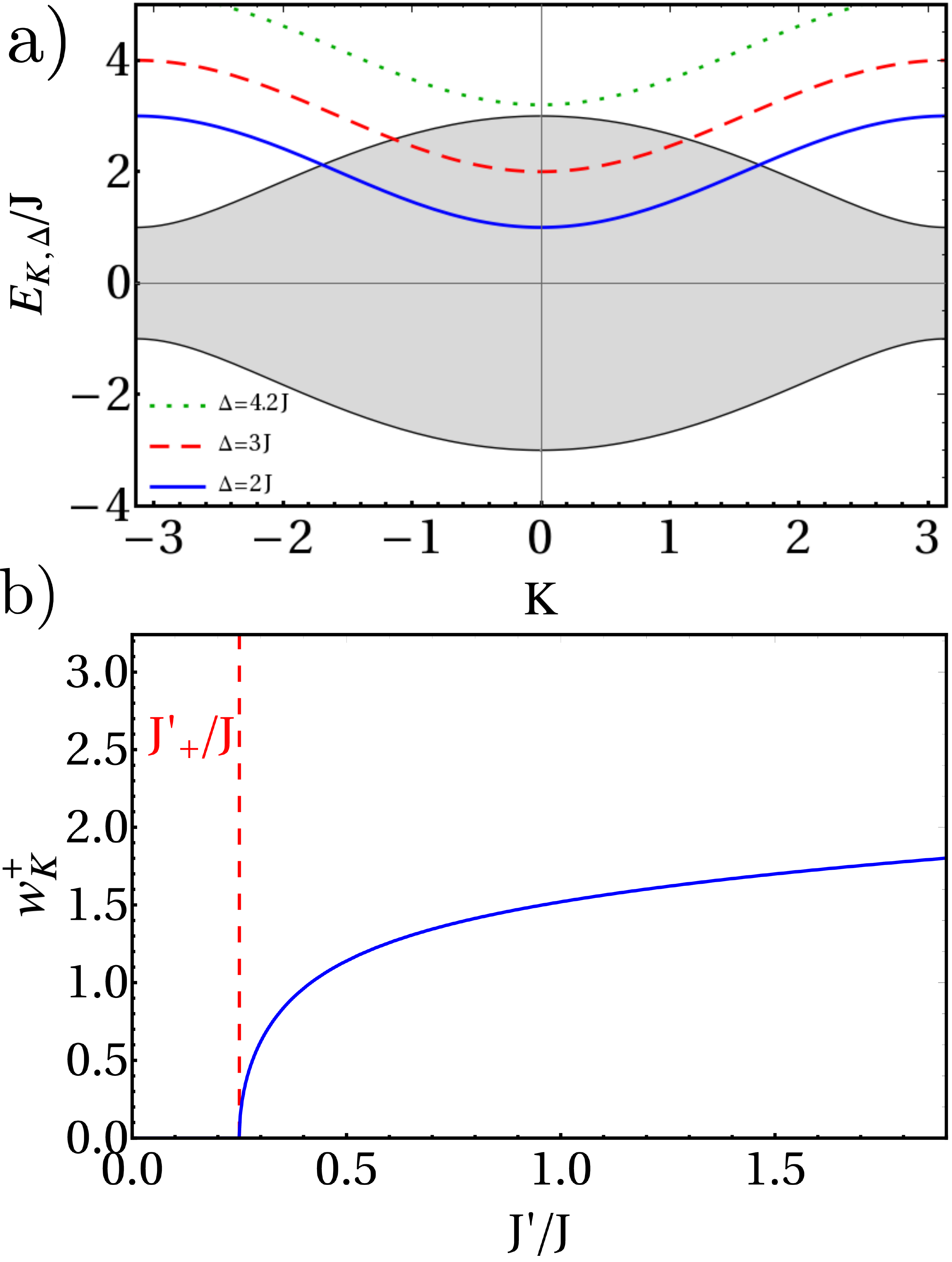}
\caption{a) Energy of the effective qubit $E_{K,\Delta}$, for $J'=0.5J$ and $\Delta=4.2J,3J,2J$ (dotted green, dashed red, and solid blue, respectively), as a function of $K$. The shaded region renders the effective band $\tilde{\omega}_{K,p}$. b) Width of the window of momenta for which the qubit undergoes spontaneous emission (see main text for details) as a function of $J/J'$, for $\Delta=3J$. The red dashed line indicates the critical value $J_+'\approx 0.25J$.
Notice that $w_K^+$ is defined as the width of the window for positive momentum $K$; there is an equivalent window for negative values of $K$.}
\label{fig:xi_up}
\end{figure}

A similar behavior is observed when $\Delta$ is close to the maximum of the photonic band $\omega_p$, i.e., when $\Delta \approx 2J$ (see Fig.~\ref{fig:xi_up}). In such case, however, the window of wavevectors $K$ for which the qubit undergoes spontaneous emission is centered around $K=0$ (see Fig.~\ref{fig:xi_up}a)). This difference stems from the asymmetry of the qubit effective energy band $E_{K,\Delta}$ with respect to its median value $\Delta$. Following a similar reasoning as above, we conclude that selective emission above the upper band edge requires $\Delta<4J'+2J$, thus defining another critical value for $J'$, namely $J_+' = \Delta/4-J/2$. The width of momenta for which spontaneous emission occurs, $w_K^+$, is displayed in Fig.~\ref{fig:xi_up}b) as a function of $J'/J$. 

\begin{figure}[tbh!]
\includegraphics[width=\linewidth]{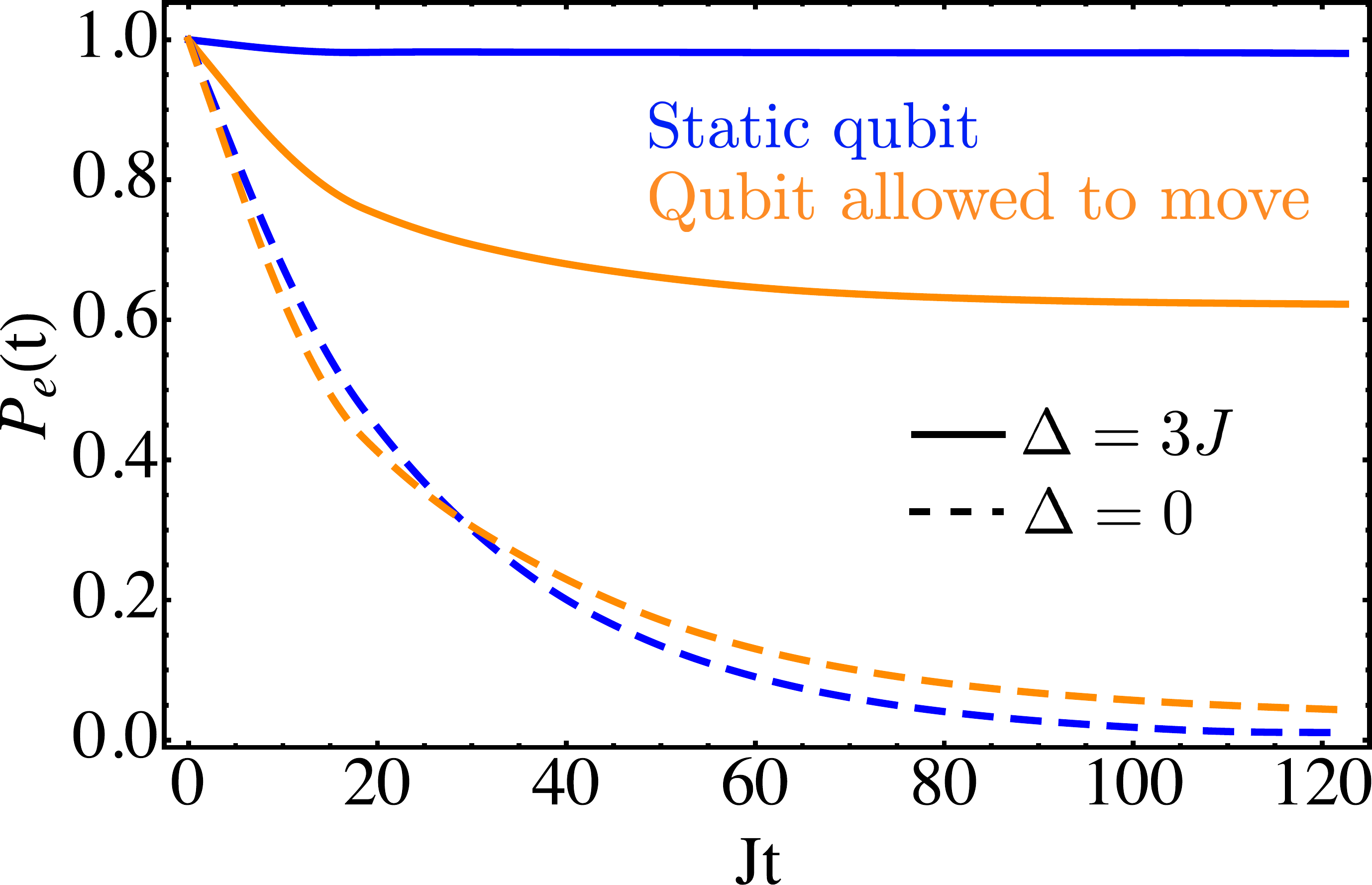}
\caption{Total qubit occupation for the static ($J'=0$, blue) and movable qubit ($J'=0.5J$, orange), for a qubit initially at $x=0$. The solid and dashed lines correspond, respectively, to a qubit energy partially embedded and fully embedded in the photonic band, i.e., $\Delta=3J$ and $\Delta=0$, respectively. }
\label{fig:pe_t}
\end{figure}

The selective emission of some $K-$ components has an impact in the usual spontaneous emission scenario, as we show in Fig. \ref{fig:pe_t}. Here, we consider the qubit initially at a well-defined position $x_0=0$ by fixing $\varphi_K=L^{-1/2}$, solve the full dynamics numerically, and compute the total excited state probability, $P_e(t) \equiv \sum_K \langle\Psi(t)\vert e^\dagger_Ke_K \vert\Psi(t)\rangle$. Figure \ref{fig:pe_t} shows this probability for a fixed qubit ($J'=0$, blue lines) and for a qubit allowed to move ($J'=0.5$, red lines), whereas the solid and dashed lines correspond, respectively, to qubit frequencies $\Delta=3J$ and $\Delta=0$. For a static qubit, these two values lie outside and embedded in the band, respectively, resulting in either a full suppression of the decay for $\Delta=3J$ and an exponential decay for $\Delta=0$. This exponential decay at $\Delta=0$ remains qualitatively unchanged for a qubit allowed to move, as the effective energies $E_{K,\Delta}$ are embedded in the effective photonic band for all $K$. However, for $\Delta=3J$, only the momenta $K$ within the emission window $w_K^+$ are embedded in the band and decay into the waveguide. As shown by Fig. \ref{fig:xi_up}b), these momenta amount to a fraction $\approx 1.1/\pi \approx 35\%$ of the total interval $K\in [-\pi,\pi]$ and, as a consequence, about $35\%$ of the initial occupation of the qubit is emitted in the form of waveguide photons. Conversely, about $65\%$ of the initial occupation remains on the qubit even at large times.

% As shown by the above results, the $K-$selective emission can have a significant impact on the spontaneous emission dynamics, which might be harnessed in the future for photonic state engineering. For instance, at parameter regimes for which $w_K^\pm$ becomes very small, the spectrum of the emitted photons might become much narrower than the qubit natural linewidth, thus potentially allowing, in the strong coupling regime, for very monochromatic single-photon emission at a high rate (albeit with very weak intensities). A second possibility could be to use spontaneous emission in these regimes to initialize bound-state wavepackets in interesting regions of their band dispersions (see e.g. the discussion around Fig.~\ref{fig:bs_energies}). Although this potential lies beyond the scope of our work, our results highlight the strong impact that qubit motion can have on the final qubit occupations.

As shown by the above results, the $K-$selective emission can have a significant impact on the spontaneous emission dynamics. One can envision that, in future works, this might be harnessed for photonic state engineering. An interesting prospect, for instance, would be to analize the feasibility of devising spectrally narrow photonic wavepackets via spontaneous emission at parameter regimes where $w_K^\pm$ becomes very small.

\subsection{Spatial distribution of photons emitted inside the band}\label{subsecSpatial}

\begin{figure}[tbh!]
\includegraphics[width=\linewidth]{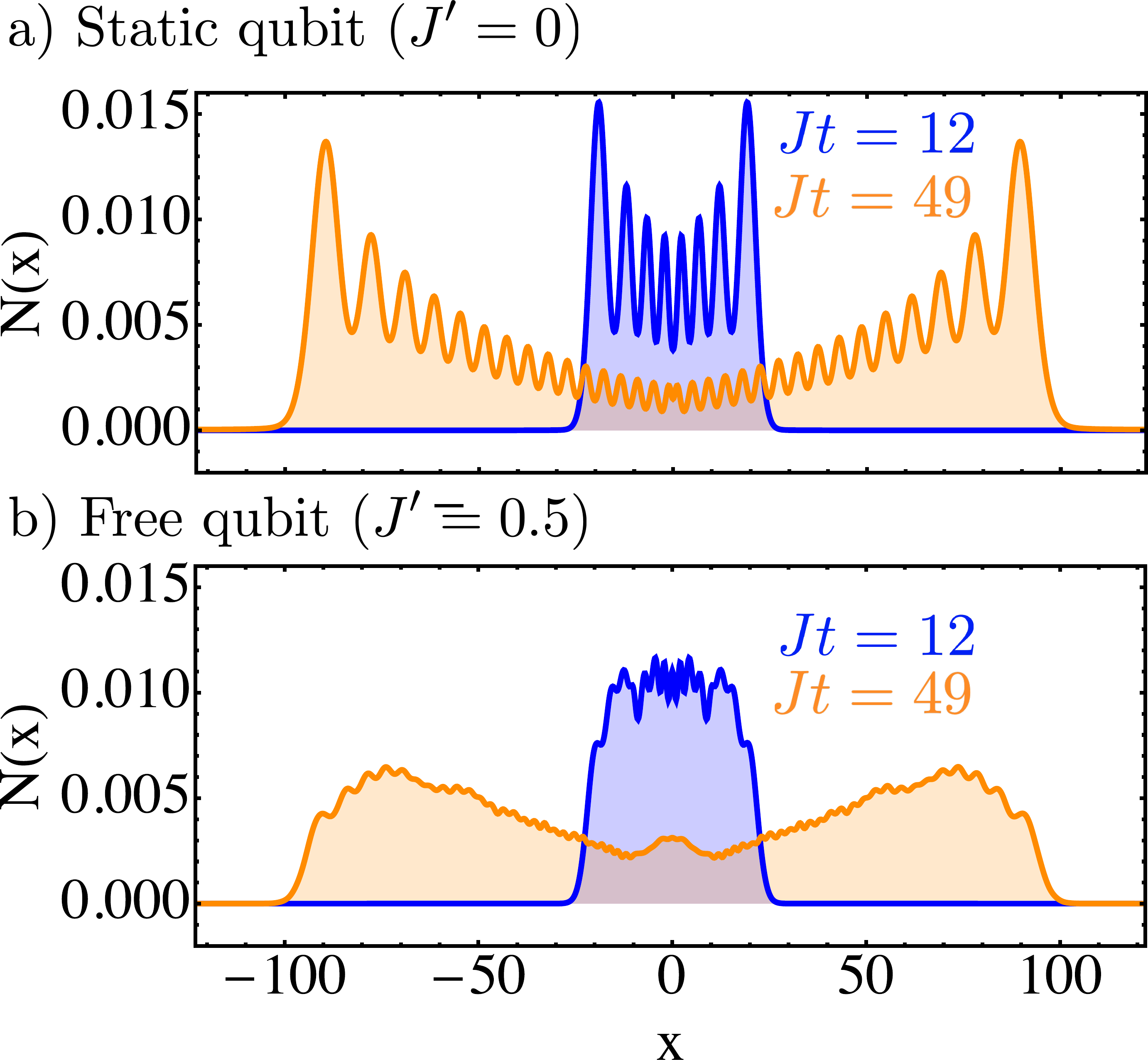}
\caption{Photonic occupation as a function of position (Eq. \ref{Nxdef}) for $\Delta=0$, $\Omega=0.2J$, and two instants of time, namely $Jt=12$ (blue lines) and $Jt=49$ (orange lines). a) Static qubit case, $J'=0$. b) Moving qubit ($J'=0.5J$).}
\label{fig:Nx}
\end{figure}

Let us briefly discuss the dynamics of the emitted photons in real space, for an initially excited qubit localized at position $x_0=0$. We obtain the full quantum state $\vert \Psi(t)\rangle$ by numerically solving the full system dynamics. We then compute the mean photonic occupation number at position $x$, namely
\begin{equation}\label{Nxdef}
    N(x,t)\equiv \frac{1}{L}\sum_{p,p'}e^{ix(p-p')}\langle \Psi(t) \vert a_p^\dagger a_{p'}\vert \Psi(t)\rangle.
\end{equation}
In Fig.~\ref{fig:Nx} we plot this occupation as a function of $x$ and at two instants of time, for $\Delta=0$, $\Omega=0.2J$, and for both a fixed qubit ($J'=0$, panel a) and a qubit allowed to move ($J'=0.5J$, panel b). The photonic distribution of Fig.~\ref{fig:Nx}a) describes free propagation of a wavepacket in a cosine dispersion relation, and recovers the well-known results for a static qubit \cite{Lombardo2014,Sanchez-Burillo2017}. For a moving qubit (Fig.~\ref{fig:Nx}b)), the emitted wavepacket keeps the global features from the static case, such as the velocity of the photonic wavefront propagation at $v_{\rm max}^{\rm qb} =  2J$ (see Eq.~\ref{eq:vph}). However, the emitted wavepacket displays significant differences with respect to the static qubit. Specifically, the spatial coherence of the wavepacket, namely the oscillatory behavior present in the static qubit case, is strongly suppressed. This is a signature of destructive quantum interference between the multiple processes leading to the detection of a photon at position $x_d$ and time $t_d$: on the one hand, the emission of the photon with a velocity $v = x_d/t_d$ from a qubit at $\{x,t\}=0$, which is the only possible process for a static qubit; on the other hand, the propagation of the excited qubit from $\{x,t\}=\{0,0\}$ to $\{x,t\}=\{x',t'\}$, followed by the
emission of a photon at a velocity $v' = (x_d-x')/(t_d-t')$. All these processes result in emission of photons at different velocities (i.e. different frequencies) that incoherently interfere at every position in the waveguide. Note that, for values of $x$ very close to the wavefront, the oscillations are still appreciable, as only a few processes result in photons at such positions. Indeed, photons detected at positions close to the wavefront are emitted both with well defined momentum $p$ (i.e. the momentum at which $v^{\rm ph} = v^{\rm ph}_{\rm max}$), and at times close to $t=0$ or, equivalently, at well-defined qubit positions $x\approx0$.

\begin{figure}[tbh!]
\includegraphics[width=\linewidth]{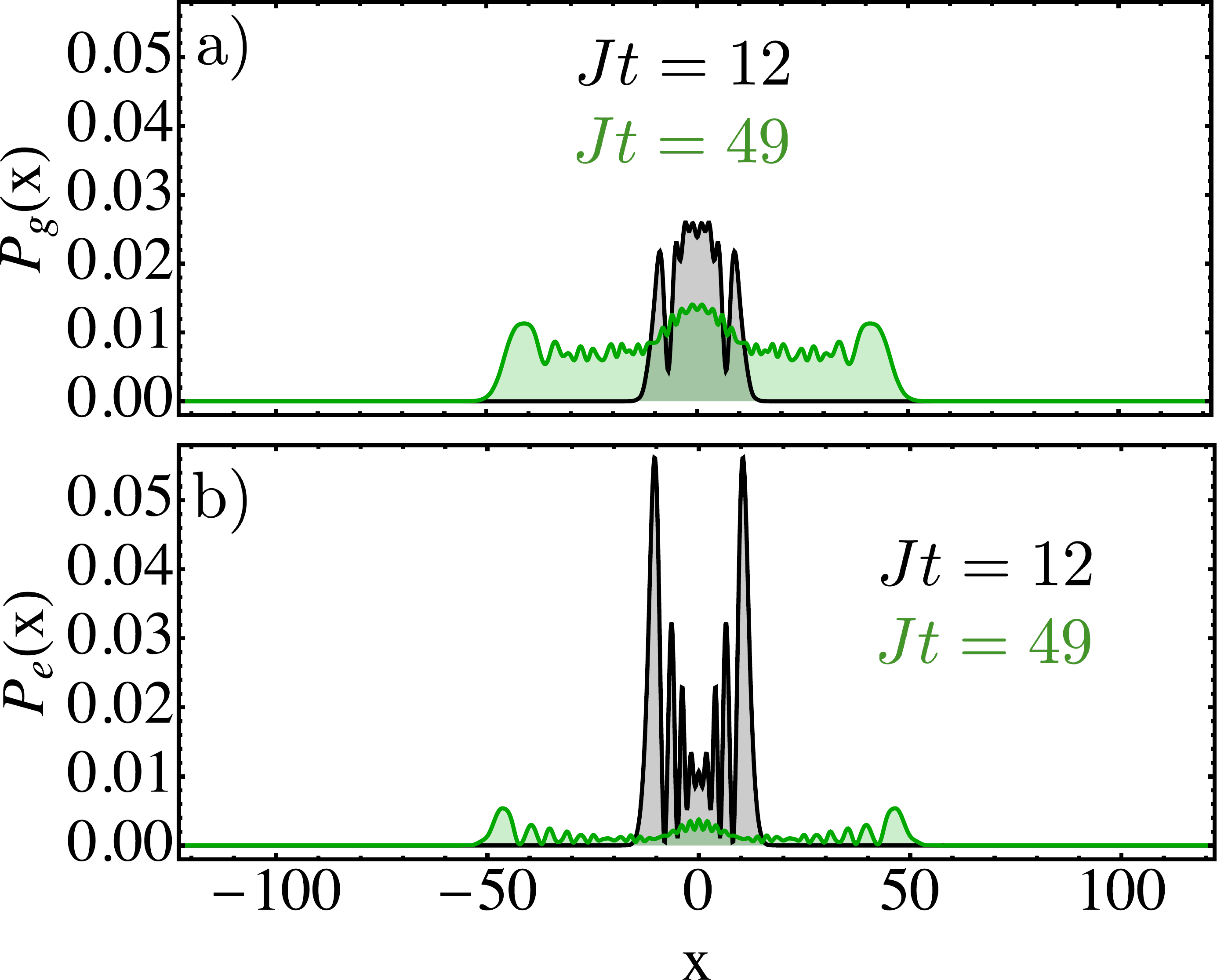}
\caption{a) Ground-state qubit occupation at position $x$ (Eq.~\eqref{Pg}) for the same parameters and instants of time as in Fig.~\ref{fig:Nx}b. b) Excited-state qubit occupation (Eq.~\eqref{Pe}) at position $x$.}
\label{fig:ExGx}
\end{figure}

We have also calculated the position distribution of the qubit both in its ground and in its excited states, given by
\begin{equation}\label{Pg}
    P_g(x,t)\equiv \frac{1}{L}\sum_{K,K'}e^{ix(K-K')}\langle \Psi(t) \vert g_K^\dagger g_{K'}\vert \Psi(t)\rangle,
\end{equation}
and
\begin{equation}\label{Pe}
    P_e(x,t)\equiv \frac{1}{L}\sum_{K,K'}e^{ix(K-K')}\langle \Psi(t) \vert e_K^\dagger e_{K'}\vert \Psi(t)\rangle,
\end{equation}
respectively. Note that for a static qubit these quantities are equal to $P_{e,g}(x,t) = P_{e,g}(t)\delta_{x}$.  In Fig. \ref{fig:ExGx} we show $P_g(x,t)$ and $P_e(x,t)$ (panels a and b, respectively) for a qubit allowed to move, i.e. for the same parameters as in Fig. \ref{fig:Nx}b. Note that since $J'=0.5J$, the wavefront for the qubit motion propagates half as fast as the photonic wavefront. Furthermore, the spatial coherence of the ground and excited states is very different: on the one hand, $P_g(x)$ shows very incoherent behavior due to a similar interference effect as described above. Indeed, many processes lead to a ground-state qubit at position $x$, namely propagation of the excited qubit until an arbitrary position followed by emission of a photon with an appropriate energy. On the other hand $P_e(x)$ shows clearly coherent oscillations, as only one process, namely propagation without emission, leads to an excited-state qubit at position $x$. Because of this, $P_e(x)$ shows the same oscillations as Fig. \ref{fig:Nx}a), corresponding to free propagation of a localized wavepacket in a cosine dispersion relation. Finally, note that in cases of $K-$selective emission, the behavior of the spatial photonic and qubit distributions can also be understood from the above arguments, as shown in Appendix \ref{AppendixAdditionalFigures}.

The results of this section evidence yet again the potentially large impact of the qubit motion on the spontaneous emission dynamics. The spatial coherence of both the motional wavepacket and the wavepacket of the emitted photons is strongly suppressed. Far from being a hindrance, this property can be harnessed in low-decoherence scenarios, where the measurement of a photon at position $x$ could herald the collapse of the motional qubit state into a large spatial superposition. Applications aside, our results emphasize the relevance of qubit motion which, as has already been done for single-photon absorption \cite{TrautmannPRA2016}, should be taken into account to appropriately model wQED systems in the quantum regime.

\section{Conclusions}\label{SecConclusions}

In this work, we have characterized the single-excitation properties of a wQED system where the qubit motion is described quantum mechanically. In this regime, accessible by various experimental setups, exotic behavior has been shown for the single-photon scattering, the bound states in the bandgap, and the spontaneous emission dynamics. Although in each section certain potential applications have been mentioned, the main results of this work are the extension of the theoretical toolbox to the moving qubit case, and the characterization of the phenomenology, especially in comparison to its static qubit counterpart. Our work paves the way to the inclusion of the qubit motion in wQED systems, if not as a resource to be harnessed, as a necessary ingredient for the accurate modeling of such platforms in the low kinetic energy regime.
Furthermore, our results uncover
 a new horizon to explore in wQED, and can be extended in multiple ways. Three particularly interesting examples are: a) Exploring cases where the lattices of the photons and the qubit have different periodicities.
 b) extending our results to multiple qubits, either to characterize their bound-state-mediated interactions in the presence of motion, or to study the effect of the motion on the collective dynamics, e.g. to characterize the bright and dark motional states of a many-qubit system; c) extending our results to the multi-photon scenario, and study , for instance, the photon-photon scattering bound states stemming from stimulated emission \cite{Zheng2010}, or the possibility of devising two-photon quantum gates based on qubit motion.

\acknowledgments

We thank Adri\'an E. Rubio L\'opez, O. Romero-Isart, Talitha Weiss, and Daniel Malz for fruitful and inspiring discussions. We are especially thankful to Prof. Tao Shi for suggesting the form of the interaction Hamiltonian \eqref{eq:Hint}, and to Giuseppe Calaj\`o for his careful reading and valuable insight. E.~S.~-B acknowledges ERC Advanced Grant QUENOCOBA under the EU Horizon 2020 program (grant agreement 742102). C.~G.~-B. acknowledges funding from the EU Horizon 2020 program under the Marie Sk\l{}odowska-Curie grant agreement no.~796725 (PWAQUTEC).  A.G.-T. acknowledges support from Project PGC2018-094792-B-I00~(MCIU/AEI/FEDER, UE), CSIC Research Platform PTI-001, and CAM/FEDER Project No.~S2018/TCS-4342~(QUITEMAD-CM). 

\appendix

\section{Self-energy}\label{app:self-energy}

\begin{figure}
    \centering
    \includegraphics[width=\linewidth]{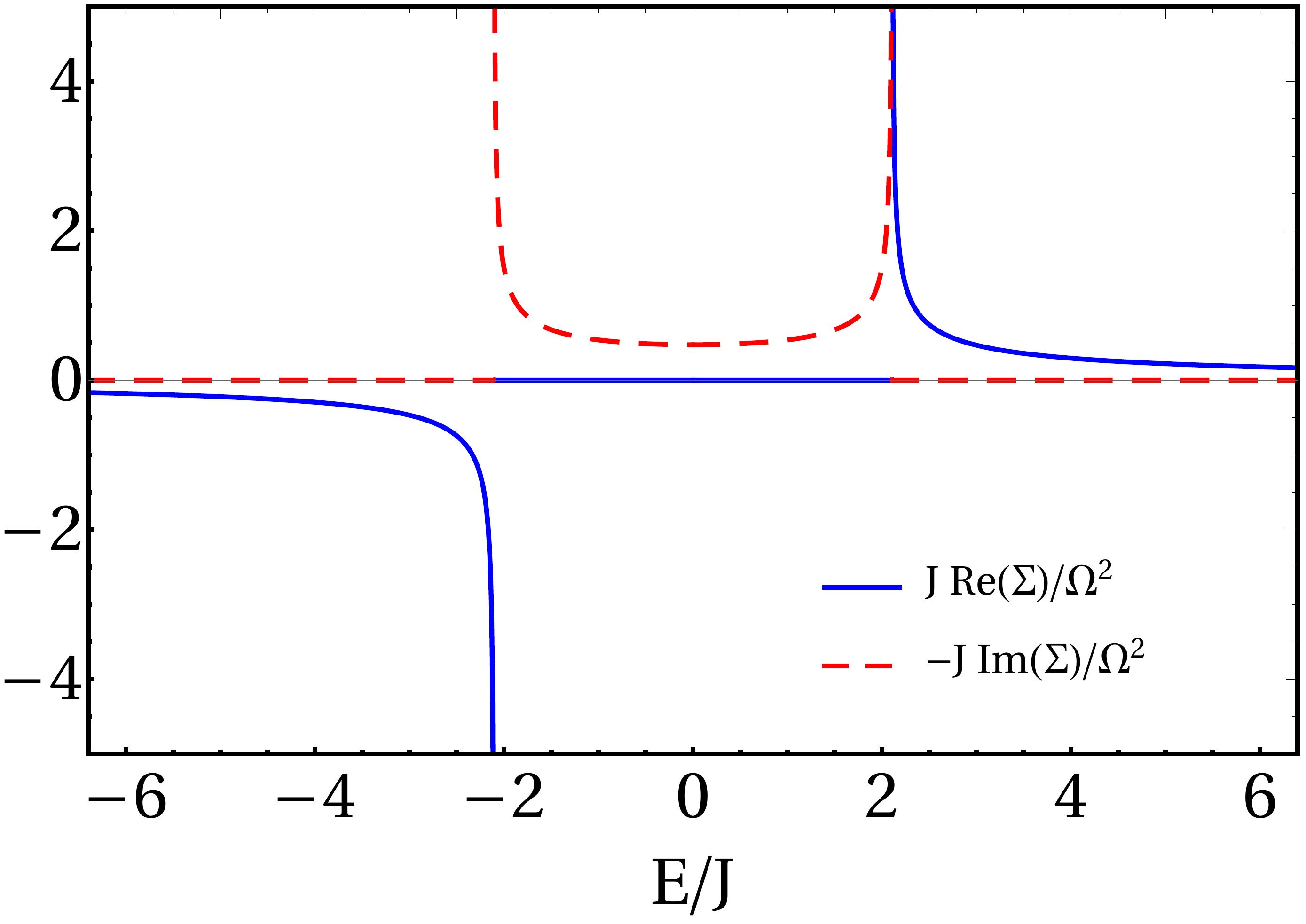}
    \caption{Real (solid blue line) and imaginary (red dashed line) parts of the self-energy in units of $\Omega^2/J$ for $J'=0.1$ and $K=\pi/3$.}
    \label{fig:self}
\end{figure}

In this section we compute the self-energy, introduced in Eq. \eqref{eq:self-energy}. We first take the change of variable $y=e^{ip}$ to write the self-energy as an integral over the complex unit circumference,
\begin{align}
    \Sigma_K(E) =& \frac{\Omega^2}{2\pi i(J+J'e^{-iK})} \nonumber \\
    & \times \rcirclerightint \frac{dy}{y^2+\frac{E}{J+J'e^{-iK}}y+\frac{J+J'e^{iK}}{J+J'e^{-iK}}}.\label{eq:self-energy_y}
\end{align}
We then find the roots of the denominator
\begin{equation}\label{eq:roots}
    y_\pm = \frac{1}{2z(K)}\left[-E\pm\sqrt{E^2-4\left\vert z(K)\right\vert^2}\right],
\end{equation}
where $z(K)$ is defined in the main text (Eq. \eqref{eq:z_def}). By adding the usual convergence factor in the physical plane, $E \to E+i0^+$, one can easily demonstrate that one and just one of the roots is inside the unit circle, namely $\vert y_+\vert<1<\vert y_-\vert$ for $E>2\vert z(K)\vert$ and $\vert y_-\vert<1<\vert y_+\vert$ for $E<2\vert z(K)\vert$. Then, applying the residue's theorem we find
\begin{equation}\label{eq:self-energy_sol}
 \Sigma_K(E) = \frac{\Omega^2}{\sqrt{E^2-4\vert z(K)\vert^2}}\text{sign}\left(E-2\vert z\vert\right)
 ,
\end{equation}
 We plot the real and imaginary part of the self-energy in Fig. \ref{fig:self}. There we see that it behaves similarly to the static case.

\section{Single-excitation $S$-matrix}\label{app:s_matrix}

In this appendix we summarize the derivation of the $S-$matrix \eqref{eq:Sfi} and, in Sec. \ref{secTR}, we detail the computation of the scattering coefficients. For simplicity of notation we assume $\Omega_p=\Omega$, but the derivation for a momentum-dependent coupling is analogous. First of all, it is convenient to express the time-evolution operator in powers of the interaction Hamiltonian \cite{fetter2003quantum}, and remove the time-ordering by expressing each term of the series as a set of nested integrals, i.e.,
\begin{equation}\label{timeevol}
\begin{split}
U(t,t_0) &= \mathbb{I} -i\int_{t_0}^t dt_1 H_{1,K}(t_1) + 
\\
&
+(-i)^2 \int_{t_0}^t dt_1\int_{t_0}^{t_1} dt_2 H_{1,K}(t_1) H_{1,K}(t_2) + ...
\end{split}
\end{equation}
with $H_{1,K}(t)$ the Hamiltonian \eqref{eq:H1K} in the interaction picture,
\begin{equation}\label{eq:Hintpicture}
\begin{split}
    H_{1,K}(t) &= e^{i(H_{1,K})_\text{bare}t} H_{1,K} e^{-i (H_{1,K})_\text{bare}t}-(H_{1,K})_\text{bare} =
    \\
    &
    =e^{i (H_{1,K})_\text{bare}t} (H_{1,K})_\text{int}e^{-i (H_{1,K})_\text{bare}t}=
    \\
    &
    = \frac{1}{\sqrt{L}}\sum_{p} \Omega e^{-i\Delta_{K,p}t} \ket{p}_{K+p}\bra{K+p}  + \text{H.c.},
\end{split}
\end{equation}
with $(H_{1,K})_\text{bare}=\sum_p \tilde{\omega}_{K,p} \ket{p}_K \bra{p}_K + E_{K,\Delta}\ket{K}\bra{K}$, and
\begin{equation}
    \Delta_{K,p} \equiv \omega_p+\xi_K-\Delta-\xi_{K+p},
\end{equation}
that is, $\Delta_{K,p}$ represents the detuning between (i) the state formed by a photon with momentum $p$ and a unexcited qubit with momentum $K$ and (ii) a state with no photons and an qubit with momentum $K+p$ in its internal excited state.

Using the expression \eqref{timeevol} for the time evolution operator, and noting that, since the interaction Hamiltonian \eqref{eq:Hint} contains single raising and lowering qubit operators, only the even terms in the series result in a non-zero contribution, we can cast the S-matrix in the form
\begin{equation}
    S_{fi} = \Big(\delta_{p_f,p_i}\delta_{k_f,k_i} + T_{fi}\Big).
\end{equation}
The $T-$matrix is defined as
\begin{equation}
    T_{fi} = \sum_{n=1}^\infty T_{2n}\propto \delta_{p_f + k_f,p_i+k_i},
\end{equation}
with
\begin{equation}
\begin{split}
    T_{2n} &= (-i)^{2n}\int_{-\infty}^\infty dt_1 \int_{-\infty}^{t_1} dt_2 ... \int_{-\infty}^{t_{2n-1}}dt_{2n} 
    \\
    &
    \bra{f} H_{1,K}(t_1)... H_{1,K}(t_{2n}) \ket{i}.
\end{split} 
\end{equation}
The $T-$matrix can be calculated exactly. First, we use Eq. \eqref{eq:Hintpicture} to construct the following building block
\begin{equation}
\begin{split}
    H(t_a)&H(t_b)\ket{i} =  H(t_a)H(t_b)\ket{p_i}_{k_i+p_i} = \frac{1}{L}\Omega e^{-i\Delta_i t_b}
    \\&
    \times\sum_p\Omega^* e^{i\Delta_{k_i+p_i-p,p}t_a}\vert p,k_i+p_i-p\rangle_g
\end{split}
\end{equation}
where we define for simplicity
\begin{equation}
    \Delta_{i} \equiv \Delta_{k_i,p_i},
\end{equation}
analogously, we define $\Delta_f$. By repeated application of the above identity, and acting on the left with the bra $\bra{f}={}_{k_f+p_f}\bra{p_f}$,
%it is straightforward to prove that
%\begin{equation}\label{buildingblock2}
%\begin{split}
%    H&(t_1)...H(t_{2n})\ket{i} = \frac{1}{L}\Omega e^{-i\Delta_i t_{2n}}
%     \\&
%     \times G_0(t_{2n-1}-t_{2n-2})...G_0(t_3-t_2)
%     \\&
%     \times\sum_p\Omega^* e^{i\Delta_{k_i+p_i-p,p}t_1}\vert p\rangle_{k_i+p_i},
% \end{split} 
% \end{equation}
% where we have defined the function
% \begin{equation}\label{G0def}
%     G_0(\tau) \equiv \frac{1}{L}\sum_p \Omega^2e^{i\Delta_{k_i+p_i-p,p}\tau}.
% \end{equation}
% The second step is to act with the bra $\bra{f}={}_{k_f+p_f}\ket{p_f}$ in the left of the above expression, obtaining
we obtain
\begin{equation}\label{buildingblock3}
\begin{split}
    \bra{f} & H(t_1)... H(t_{2n})\ket{i} =
    \\
    &
    =\delta_{p_f + k_f,p_i+k_i}\frac{1}{L} \vert \Omega \vert^2 e^{-i\Delta_i t_{2n}}e^{i\Delta_ft_1}
    \\
    &
    \times G_0(t_{2n-1}-t_{2n-2})...G_0(t_3-t_2),
\end{split}
\end{equation}
where we have defined the function
\begin{equation}\label{G0def}
    G_0(\tau) \equiv \frac{1}{L}\sum_p \Omega^2e^{i\Delta_{k_i+p_i-p,p}\tau}.
\end{equation}
Note that the conservation of total momentum, $\delta_{p_f + k_f,p_i+k_i}$, arises naturally from the calculation. Using the above identity we can construct an inductive relation between the coefficients $T_{2n}$. We do this by constructing explicitly the term $2n+2$, and rearranging the terms as
\begin{widetext}
\begin{equation}
\begin{split}
    T_{2n+2} &= (-i)^{2n}\int_{-\infty}^\infty dt_1 \int_{-\infty}^{t_1} dt_2 ... \int_{-\infty}^{t_{2n-1}}dt_{2n} 
    \delta_{p_f+k_f,p_i+k_i} \frac{\Omega^2}{L}e^{i\Delta_ft_1} G_0(t_{2n-1}-t_{2n-2})...G_0(t_3-t_2)
    \\
    &
    \times (-i)^2 \int_{-\infty}^{t_{2n}}dt_{2n+1}\int_{-\infty}^{t_{2n+1}}dt_{2n+2} e^{-i\Delta_i t_{2n+2}}G_0(t_{2n+1}-t_{2n})
\end{split}
\end{equation}
\end{widetext}
We can now explicitly perform the integrations in the last line using Eq. \eqref{G0def}, to find
\begin{equation}
\begin{split}
    \int_{-\infty}^{t_{2n}}&\!\!dt_{2n+1}\int_{-\infty}^{t_{2n+1}}\!\!dt_{2n+2} e^{-i\Delta_i t_{2n+2}}G_0(t_{2n+1}-t_{2n})=
    \\
    &
    =
    \frac{e^{-i\Delta_i t_{2n}}}{L}\sum_p\frac{\Omega^2}{\Delta_i(\Delta_{k_i+p_i-p,p}-\Delta_i)} \equiv e^{-i\Delta_i t_{2n}} K_0.
\end{split}
\end{equation}
To perform the integral, we have included a convergence factor $\exp(-\epsilon \vert t_j\vert)$ in each integral following the ``adiabatic switching-on'' argument~\cite{fetter2003quantum}.
Combining the above two expressions we find the recurrence relation
\begin{equation}
    T_{2n+2} = -K_0T_{2n},
\end{equation}
from where the $T-$matrix can be immediately written as
\begin{equation}
    T_{fi} = T_2(1-K_0+K_0^2-...) = \frac{T_2}{1+K_0}.
\end{equation}

As a final step, we calculate the coefficient $T_2$ using Eq. \eqref{buildingblock3}:
\begin{equation}
\begin{split}
    T_2 &= \delta_{p_f + k_f,p_i+k_i}\frac{-1}{L}\vert \Omega\vert^2
    \\
    &\int_{-\infty}^\infty dt_1 e^{i\Delta_ft_1} \int_{-\infty}^{t_1}dt_2e^{-i\Delta_i t_{2}}=
    \\
    &
    =\delta_{p_f + k_f,p_i+k_i}\frac{-2\pi i}{L}\Omega^2 \frac{\delta(\Delta_f-\Delta_i)}{\Delta_i},
\end{split}
\end{equation}
where the last Dirac delta enforces energy conservation. The total $S-$matrix can thus be written as
\begin{equation}\label{Smatrixdiscrete}
\begin{split}
    S_{fi} &= \delta_{p_f + k_f,p_i+k_i}\bigg[\delta_{p_f,p_i}
    \\
    &-\delta(\Delta_f-\Delta_i) \frac{(2\pi i/L)\vert\Omega\vert^2}{\Delta_i - \frac{1}{L}\sum_p\frac{\Omega^2}{\Delta_i}-\Delta_{k_i+p_i-p,p}}\bigg],
\end{split}
\end{equation}
or, in the continuum limit,
\begin{equation}\label{Smatrixcontinuum}
\begin{split}
    S_{fi} &= \delta(p_f + k_f-p_i-k_i)\bigg[\delta(p_f-p_i) \\
    &-\delta(\Delta_f-\Delta_i)i \Omega^2\frac{1}{\Delta_i - \frac{1}{2\pi}\int dp\frac{\Omega^2}{\Delta_i-\Delta_{k_i+p_i-p,p}}}\bigg].
\end{split}
\end{equation}

\subsection{Computing the transmission coefficients}\label{secTR}

In order to compute the scattering coefficients, we first aim at calculating the integral in the denominator of the $S-$matrix Eq. \eqref{Smatrixcontinuum}:
\begin{equation}\label{eq:self_energy}
\begin{split}
    & \frac{\Omega^2}{2\pi} \int dp \frac{1}{\Delta_i-\Delta_{k_i+p_i-p,p}}
    \\
    & =\frac{\Omega^2}{2\pi}\int dp \frac{1}{\tilde{\omega}_{k_i+p_i,p_i}-\tilde{\omega}_{k_i+p_i,p}}=\Sigma_{k_i+p_i}(\tilde{\omega}_{k_i+p_i,p_i}).
\end{split}
\end{equation}
This quantity is, by definition, the self-energy Eq. \eqref{eq:self-energy} for $K=k_i+p_i$ evaluated at $E=\tilde{\omega}_{k_i+p_i,p_i}$. Using \eqref{eq:self-energy_sol} and the fact that $\vert \tilde{\omega}_{k_i+p_i,p_i}\vert < 2\vert z(k_i+p_i)\vert $ we write
\begin{equation}
    \Sigma_{k_i+p_i}(\tilde{\omega}_{k_i+p_i,p_i}) = \frac{i\Omega^2}{\sqrt{4\vert z(k_i+p_i)\vert^2-E_i^2}}.
\end{equation}
One can also write the above equation in terms of the initial momenta $p_i$ and $k_i$ by using the equality
\begin{equation}
     4\vert z(k_i+p_i) \vert^2 \tilde{\omega}_{k_i+p_i,p_i}^2
    =
    4\left[J\sin p_i - J'\sin k_i\right]^2 ,
\end{equation}
so
\begin{equation}\label{eq:Sigmascattering}
    \Sigma_{k_i+p_i}(\tilde{\omega}_{k_i+p_i,p_i}) = \frac{i\Omega^2}{2\left[J\sin p_i - J'\sin k_i\right]}.
\end{equation}
We finally define
\begin{equation}
    \Gamma_i \equiv |\Sigma_{k_i+p_i}(\tilde{\omega}_{k_i+p_i,p_i})| = \frac{\Omega^2}{2\left[J\sin p_i - J'\sin k_i\right]}.
\end{equation}
Introducing this in the $S-$matrix \eqref{Smatrixcontinuum}
\begin{equation}
\begin{split}
    S_{fi} &= \delta(p_f+k_f-p_i-k_i)
    \\
    &
    \times \bigg[\delta(p_f-p_i)-\delta(\Delta_f-\Delta_i)\frac{i\Omega^2}{\Delta_i-i \Gamma_i}\bigg]
\end{split}
\end{equation}

The second thing we need to compute the $S$-matrix coefficients is the allowed final wavevectors, obtained from the Dirac deltas in the S-matrix. These wavevectors are thus given by conservation of momentum,
\begin{equation}
    k_f + p _f = k_i + p_i,
\end{equation}
and the conservation of energy,
\begin{equation}
    \tilde{\omega}_{k_f+p_f,p_f} = \tilde{\omega}_{k_i+p_i,p_i}.
\end{equation}
Note that the solution $p_{f,1} = p_i$ trivially fulfills the above equations, corresponding to the transmitted component. To obtain the remaining solutions, we introduce above the expressions of $\omega_p$ and $\xi_k$ (see Eqs. \eqref{eq:wp} and \eqref{eq:xik}) we find the equation
\begin{equation}\label{eqforpf}
    -\frac{\tilde{\omega}_{k_i+p_i,p_i}}{2} =  (J + J'\cos (k_i+p_i)) \cos p_f + J'\sin (k_i+p_i) \sin p_f.
\end{equation}
Naming $X\equiv \cos p_f$ we can recast the above equation as
% \begin{equation}
% \begin{split}
%     & \frac{\tilde{\omega}_{k_i+p_i,p_i}^2}{4} + X^2 (J + J'\cos(k_i+p_i))^2 \\
%     & + \tilde{\omega}_{k_i+p_i,p_i} (J + J'\cos (k_i+p_i)) X
%     \\
%     &
%     =J'^2\sin^2(k_i+p_i) (1-X^2),
% \end{split}
% \end{equation}
% or equivalently,
\begin{equation}
\begin{split}
    A X^2 +BX+C=0,
\end{split}
\end{equation}
where
\begin{equation}
    A = (J^2 + J'^2 + 2JJ'\cos (k_i+p_i) ),
\end{equation}
\begin{equation}
    B=\tilde{\omega}_{k_i+p_i,p_i} (J + J'\cos (k_i+p_i)),
\end{equation}
\begin{equation}
    C=\frac{\tilde{\omega}_{k_i+p_i,p_i}^2}{4}-J'^2\sin^2(k_i+p_i).
\end{equation}
Note that by taking the square of the equation we have introduced extra solutions that will have to be discarded. Instead of solving for the above second degree equation, we use the fact that we already know one of the solutions, namely $p_f = p_i$ or, equivalently, $X_1 = \cos p_i$. Then, we can write the second degree equation as
\begin{equation}
    A(X-X_1)(X-X_2)=0,
\end{equation}
From which the second solution of the equation is obtained as
\begin{equation}
    X_2 = \frac{C}{AX_1},
\end{equation}
Because of the parity of the cosine function, the above expression allows for two solutions for the final photon momentum,
\begin{equation}
    p_f = \pm \arccos\left[\frac{\tilde{\omega}_{k_i+p_i,p_i}^2-4J'^2\sin^2(k_i+p_i)}{4\cos(p_i)(J^2 + J'^2 + 2JJ'\cos (k_i+p_i) )}\right],
\end{equation}
only one of which is a true solution of the original equation \eqref{eqforpf}. To obtain the sign, we use the definition of $\tilde{\omega}_{k_i+p_i,p_i}$ and write the above equation only in terms of $p_i$ and $k_i+p_i$ for convenience. We then introduce the resulting expression into Eq. \eqref{eqforpf} and, after lengthy but straightforward algebra, we determine the correct sign for each given initial conditions, obtaining the final expression Eq. \eqref{eq:pf2} in the main text in.
%\onecolumngrid
%\begin{center}
%\begin{figure}
% 	\centering
% 	\includegraphics[width=\linewidth]{panels.pdf}
% 	%\vspace{-0.2cm}
% 	\caption{(Color online). a-c) Solution for the final photonic momentum $p_f$ versus initial photon and qubit momenta, $p_i$ and $k_i$ respectively (the second solution, not shown here, is the trivial $p_f = p_i$). d-f) ``Scattering diagram'' showing the scattering process occurring for every point in panels (a-c). These diagrams differentiate between transmission ($\text{sign}(p_f) = \text{sign}(p_i)$, orange) and reflection ($\text{sign}(p_f) = \text{sign}(p_i)$, green) processes.}\label{figpf}
%\end{figure}
%\end{center}
%\twocolumngrid

We are finally in a position to compute the scattering coefficients. Since, as shown by the above derivation, conservation of energy and momentum allows for only two independent final photon momenta, namely $p_{f,1} = p_i$ and $p_{f,2}$, we can write the Dirac delta in the $S$-matrix Eq. \eqref{Smatrixcontinuum} as
\begin{equation}
\begin{split}
    \delta(\Delta_i-\Delta_f) &=\left\vert\frac{d}{dp_f}\Delta_f\right\vert^{-1}_{p_f=p_i}\delta(p_f-p_i) + 
    \\
    &
    +
    \left\vert\frac{d}{dp_f}\Delta_f\right\vert^{-1}_{p_f=p_{f,2}}\delta(p_f-p_{f,2}) 
    \\
    &
    \equiv a_t \delta(p_f-p_i) + a_r \delta(p_f-p_{f,2}).
\end{split}
\end{equation}
This, combined with the definition of the $S$-matrix, Eq. \eqref{Smatrixcontinuum}, defines the scattering coefficients as
\begin{equation}
    t(k_i,p_i) \equiv 1-a_t\frac{i\Omega^2}{\Delta_i-i\Gamma_i}
\end{equation}
\begin{equation}
    r(k_i,p_i) \equiv -a_r\frac{i\Omega^2}{\Delta_i-i\Gamma_i}.
\end{equation}
The calculation of the coefficients $a_t$and $a_r$ is straightforward. One must note that the derivative of $\Delta_f$ has to be taken under the constrain of total momentum conservation, i.e.,
\begin{equation}
\begin{split}
    \frac{d\Delta_f}{dp_f} &= \frac{d\omega_p}{\omega_p}\Big\vert_{p=p_f}-\frac{d\xi_k}{dk}\Big\vert_{k=k_i+p_i-p_f}=
    \\
    &
    = 2\sin p_f -2 J\sin(k_i+p_i-p_f).
\end{split}
\end{equation}
Evaluating in both values of the final momentum, $p_{f}=p_i$ and $p_{f,2}$, is straightforward by making use of the conserved quantities, and yields
% The evaluation of the above derivative in $p_f=p_i$ is straightforward:
% \begin{equation}
%   \begin{split}
%       \left\vert \frac{d\Delta_f}{dp_f}\right\vert_{p_f=p_i}&=2\left\vert\sin p_i - J\sin k_i\right\vert =
%       \\
%       &
%       =
%       \sqrt{4\vert z(k_i+p_i)\vert^2-\tilde{\omega}_{k_i+p_i,p_i}^2},
%   \end{split}
% \end{equation}
% For $p_f= p_{f,2}$ (Eq. \eqref{eq:pf2}), we  find
% \begin{equation}
% \begin{split}
%     \left\vert \frac{d\Delta_f}{dp_f}\right\vert_{p_f=p_{f,2}}&=2\left\vert\sin p_f - J\sin k_f\right\vert_{p_f=p_{f,2}} \\
%     &= \sqrt{4\vert z(k_i+p_i)\vert^2-\tilde{\omega}_{k_i+p_i,p_i}^2}.
% \end{split}
% \end{equation}
% From the above calculations we derive
\begin{equation}
    a_t=a_r=\frac{1}{\sqrt{4\vert z(k_i+p_i)\vert^2-\tilde{\omega}_{k_i+p_i,p_i}^2}}=\frac{\Gamma_i}{\Omega^2},
\end{equation}
which allows us to finally write the scattering coefficients as
\begin{equation}
    t(k_i,p_i) = \frac{\Delta_i}{\Delta_i+i \Gamma_i},
\end{equation}
\begin{equation}
    r(k_i,p_i) = \frac{-i\Gamma_i}{\Delta_i+i\Gamma_i}.
\end{equation}
These are the expressions given in the main text, namely Eqs. \eqref{eq:t} and \eqref{eq:r}, and they fulfill the usual relations,
\begin{equation}
    1+r(k_i,p_i)=t(k_i,p_i),
\end{equation}
\begin{equation}
    \vert t(k_i,p_i)\vert^2+\vert r(k_i,p_i)\vert^2=1.
\end{equation}
Note that these relations, essential to characterize the single-photon scattering, are not fulfilled in a parametrically driven system \cite{Calajo2017}
.

\section{Bound-state computations}\label{app:bs}

We compute here $\tilde{f}_{K,\pm}^\text{bs}(x)$, the Fourier transform of $f_{K,\pm}^\text{bs}(p)$ (Eq. \eqref{eq:fkp}):
\begin{equation}
    \tilde{f}_{K,\pm}^\text{bs}(x) = \frac{\Omega u_{K,\pm}^\text{bs}}{L}\sum_p \frac{e^{ipx}}{\mathcal{E}_{K,\pm}-\tilde{\omega}_{K,p}}.
\end{equation}
Notice that $\tilde{f}_{K,\pm}^\text{bs}(x)$ is very similar to the self-energy \eqref{eq:self-energy}, being the main difference the factor $e^{ipx}$ in the sum. We take again the thermodynamic limit ($L\to\infty$). The change of variable now depends on $x$: $y=e^{ip\, \text{sgn}(x)}$, where $\text{sgn}(x)$ is the sign function. Then
\begin{equation}\label{eq:fx}
\begin{split}
    \tilde{f}_{K,\pm}^\text{bs}(x) & = \frac{\Omega u_{K,\pm}^\text{bs}}{2\pi i(J+J'e^{-iK\text{sgn}(x)})} \\
     & \times \oint dy\frac{y^{|x|}}{y^2 + \frac{\mathcal{E}_{K,\pm}}{J+J'e^{-iK\text{sgn}(x)}} y + \frac{J+J'e^{iK\text{sgn}(x)}}{J+J'e^{-iK\text{sgn}(x)}}}.
     \end{split}
\end{equation}
The poles of the integrand are those of the self-energy (see Eq. \eqref{eq:self-energy_y}) if $x>0$ and their complex conjugates if $x<0$. Then, we trivially get Eqs. \eqref{eq:fkxp} and \eqref{eq:fkxm}.

\begin{figure}
    \centering
    \includegraphics[width=0.9\linewidth]{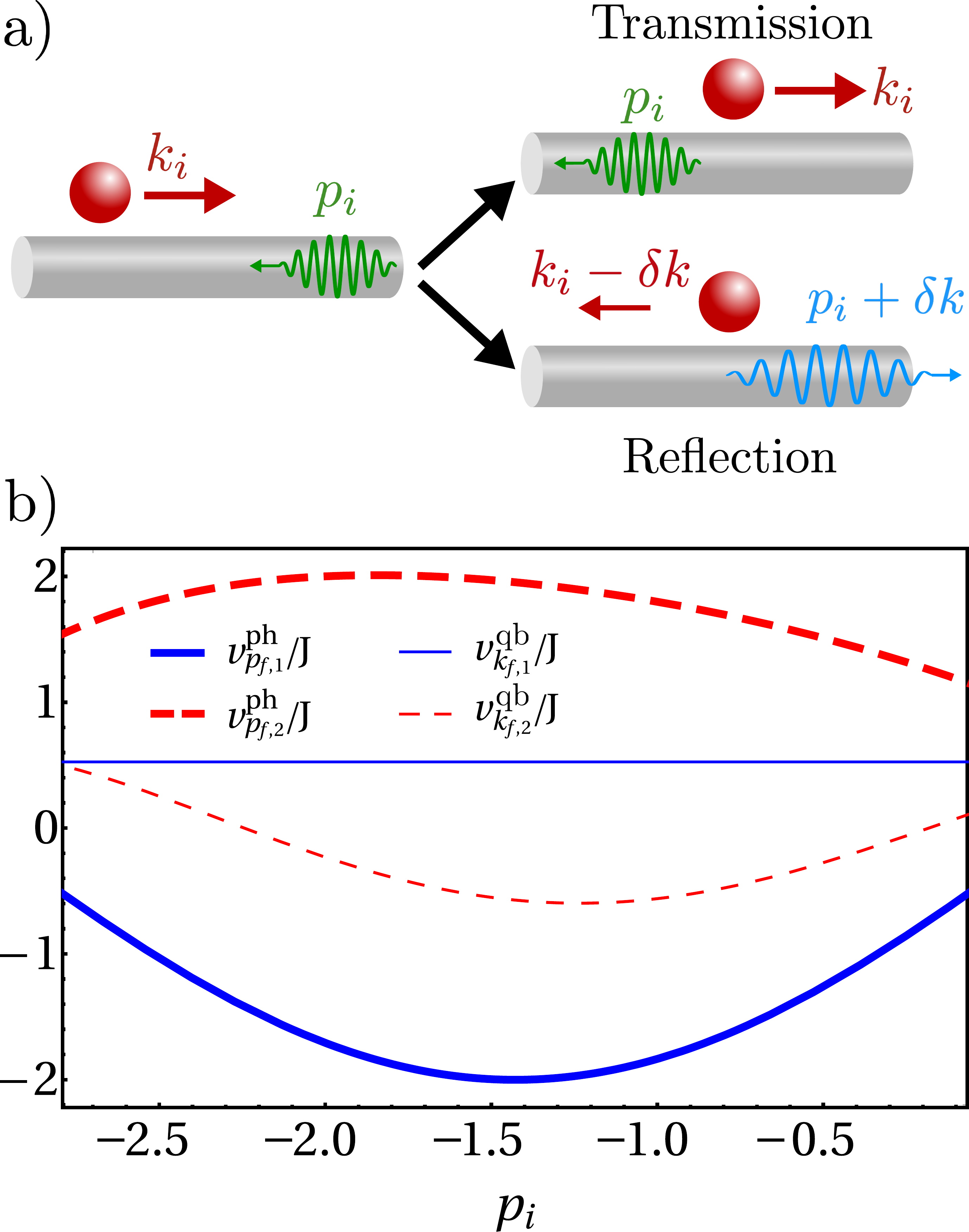}
    \caption{a) Illustration of the two allowed scattering processes in the case $\text{sign}[p_i]\ne \text{sign}[k_i]$. b) Final group velocities of photon (thick lines) and qubit (thin lines) for the transmission (blue) and reflection (red) processes. Here we choose $J'=0.3J$, $k_i=\pi/3$, and for simplicity restrict the plot to initial momenta $p_i$ for which the absolute velocity of the photon is smaller than that of the qubit, i.e., 
     $\vert v_{p_i}^\text{ph}\vert>v_{k_i}^\text{qb}$. 
    }
    \label{fig:v_appendix}
\end{figure}

\begin{figure}[tbh!]
\includegraphics[width=\linewidth]{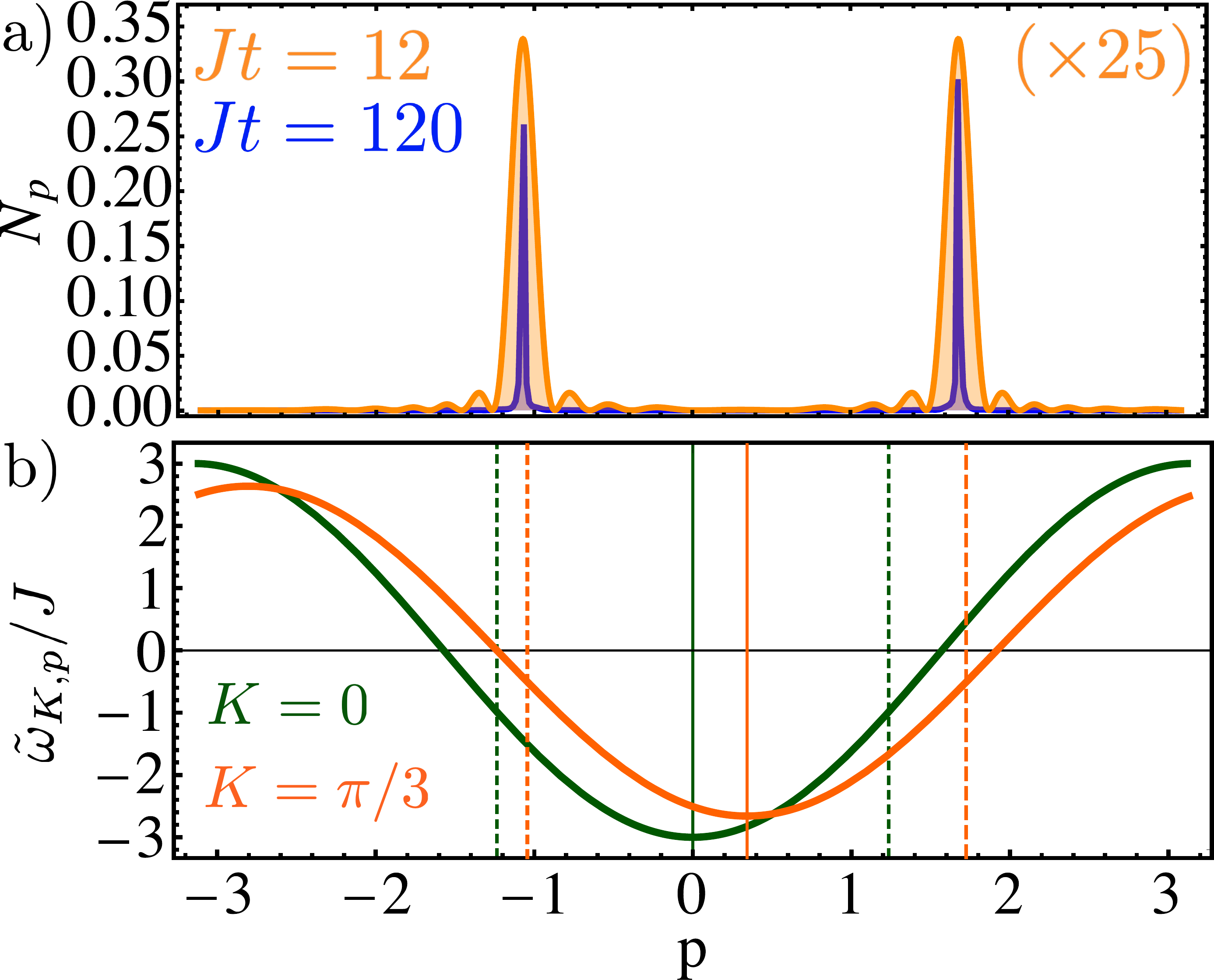}
\caption{a) Occupation of photon mode $p$ at short and long times (orange and blue lines, respectively), for $K=\pi/3$, $J'=0.5J$, $\Omega=0.2J$, and $\Delta=0$. The solutions of Eq. \eqref{eq:ppm} for these parameters are $p_+ \approx  -1.05$ and $p_-\approx 1.71$. b) Effective photonic band, $\tilde{\omega}_{K,p}$ for $K=0$ (dark green line) and $K=\pi/3$ (orange line). The vertical solid grid lines mark the minimum of each effective band, and the dashed grid lines mark the values of $p_\pm$ for each case.}
\label{fig:emission_K_appendix}
\end{figure}

\section{Spontaneous emission. Perturbative analysis}\label{app:perturbative}

Finding exact expressions for the different regimes we described in the spontaneous emission section, Sect. \ref{sect:se}, is not trivial. We therefore consider here the regime where $J'$ is small enough compared to $J$ in order to get some insight on the phenomenology.

Let us find the band limits as a function of $K$. The extreme points, which we denote as $\bar{p}$, are:
\begin{equation}\label{eq:pcrit}
    \left.\frac{\partial \tilde{\omega}_{K,p}}{\partial p}\right|_{p=\bar{p}}=0 \Rightarrow \tan \bar{p} = \frac{J'\sin K}{J+J'\cos K}.
\end{equation}
Evaluating $\tilde{\omega}_{K,p}$ at $\bar{p}$ and expanding the expression up to second order in $J'$:
\begin{equation}\label{eq:wcrit}
    \tilde{\omega}_{K,\bar{p}_\pm} = \pm\left(2J + 2J' \cos K + \frac{J'^2}{J}\sin^2 K\right) + \mathcal{O}\left(\frac{J'^3}{J^2}\right).
\end{equation}
The sign is determined by the branch we choose for the tangent in Eq. \eqref{eq:pcrit}, giving the minimum, $\bar{p}_-$, or the maximum, $\bar{p}_+$, depending on the choice. Let us focus in the low part of the band, so we consider the minimum (minus sign in Eq. \eqref{eq:wcrit}). Comparing that to the qubit energy, $E_{K,\Delta}=\Delta -2J'\cos K$, it is clear that $E_{K,\Delta}>\tilde{\omega}_{K,\bar{p}_-}$ provided $\Delta>-2J$. In other words, when $\Delta>-2J$, then the qubit is embedded in $\tilde{\omega}_{K,p}$ $\forall K$, so it will spontaneously decay no matter the value of $K$.

We consider now the case $\Delta<-2J$. In the limit $J'=0$, it is clear that $E_{K,\Delta}<\tilde{\omega}_{K,\bar{p}_-}$ $\forall K$. In order to see the effects of increasing $J'$, we take the derivative of $\tilde{\omega}_{K,\bar{p}_-}$ (Eq. \eqref{eq:wcrit}) with respect to $K$
\begin{equation}
    \frac{\partial \tilde{\omega}_{K,\bar{p}_-}}{\partial K} = 2J'\sin K -2\frac{J'^2}{J}\sin K \cos K + \mathcal{O}\left(\frac{J'^3}{J^2}\right).
\end{equation}
The slope of $\tilde{\omega}_{K,\bar{p}_-}$ will be smaller (in absolute value) than that of $E_{K,\Delta}$ in $(-\pi/2,\pi/2)$ and larger otherwise. This implies that, if $E_{K,\Delta}<\tilde{\omega}_{K,\bar{p}_-}$ at $K=\pm \pi/2$, then $E_{K,\Delta}<\tilde{\omega}_{K,\bar{p}_-}$ $\forall K$, whereas if $E_{K,\Delta}>\tilde{\omega}_{K,\bar{p}_-}$ at $K=\pm \pi/2$, there are some values of $K$ around $\pm\pi/2$ where $E_{K,\Delta}>\tilde{\omega}_{K,\bar{p}_-}$. Considering then the equation $E_{K,\Delta}=\tilde{\omega}_{K,\bar{p}_-}$ at $K=\pm \pi/2$ we can find the conditions for the parameters $J'$, $\Delta$, and $J$ such that $E_{K,\Delta}$ is/is not embedded in $\tilde{\omega}_{K,p}$ for some values of $K$:
\begin{equation}
    \left(E_{K,\Delta}=\tilde{\omega}_{K,\bar{p}_-}\right)_{K=\pm\pi/2}\Rightarrow J'_- = \sqrt{-\Delta J -2J^2}.
\end{equation}
Then, if $J'<J'_-$ the qubit \emph{is never embedded} in $\tilde{\omega}_{K,p}$, so it does not decay, whereas it decays for some values of $K$ around $\pm\pi/2$ if $J'>J'_-$.

\section{Additional figures}\label{AppendixAdditionalFigures}

\begin{figure}[tbh!]
\includegraphics[width=\linewidth]{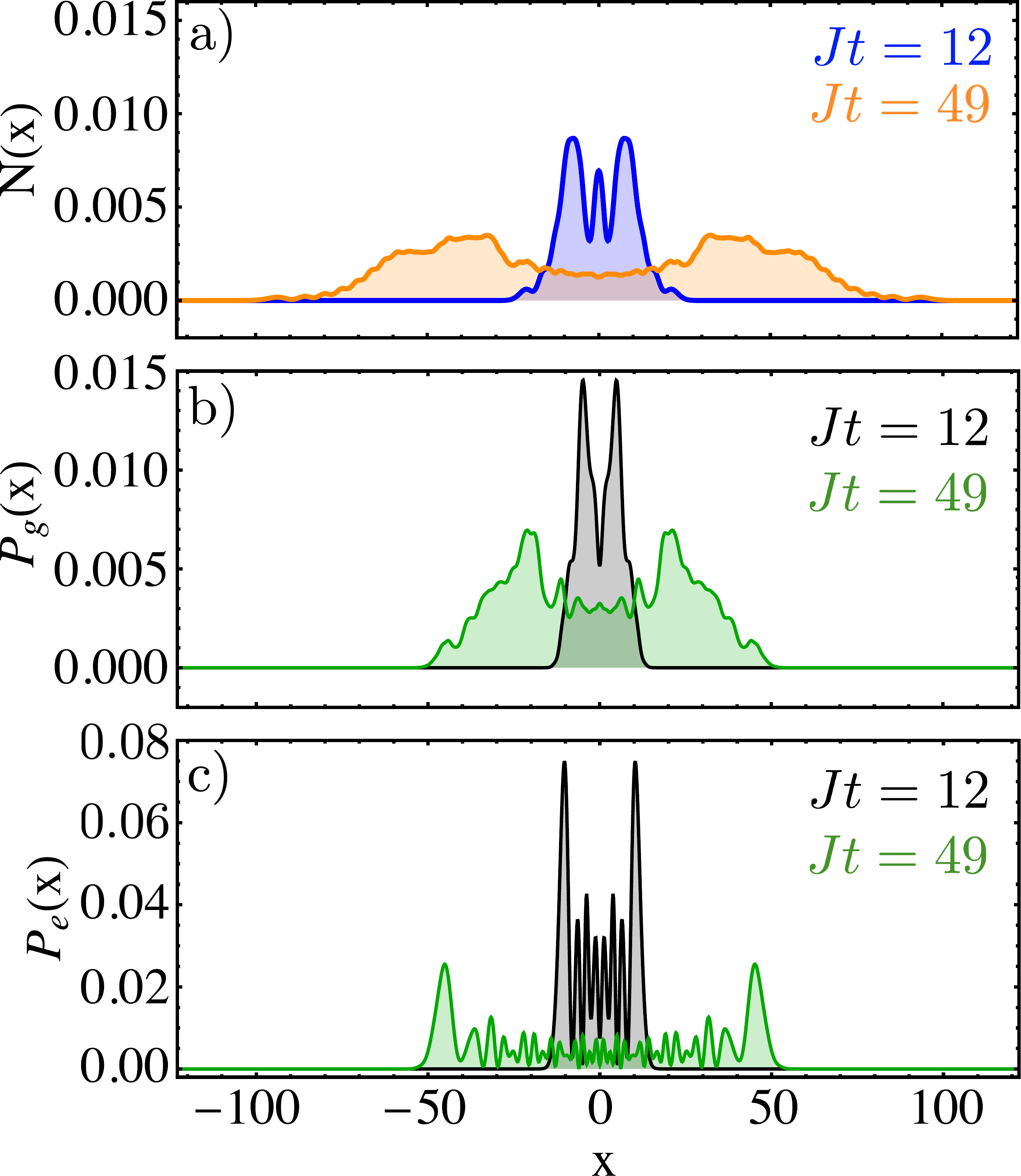}
\caption{Photonic (a), ground-state (b), and excited-state occupation (c) as a function of position, Eqs.~\eqref{Nxdef},\eqref{Pg}, and \eqref{Pe} respectively, for  $\Delta=3J$, $J'=0.5J$, $\Omega=0.2J$, and two instants of time, namely $Jt=12$ (blue/black lines) and $Jt=49$ (orange/green lines).}
\label{fig:auxDelta3}
\end{figure}

\begin{figure}[tbh!]
\includegraphics[width=\linewidth]{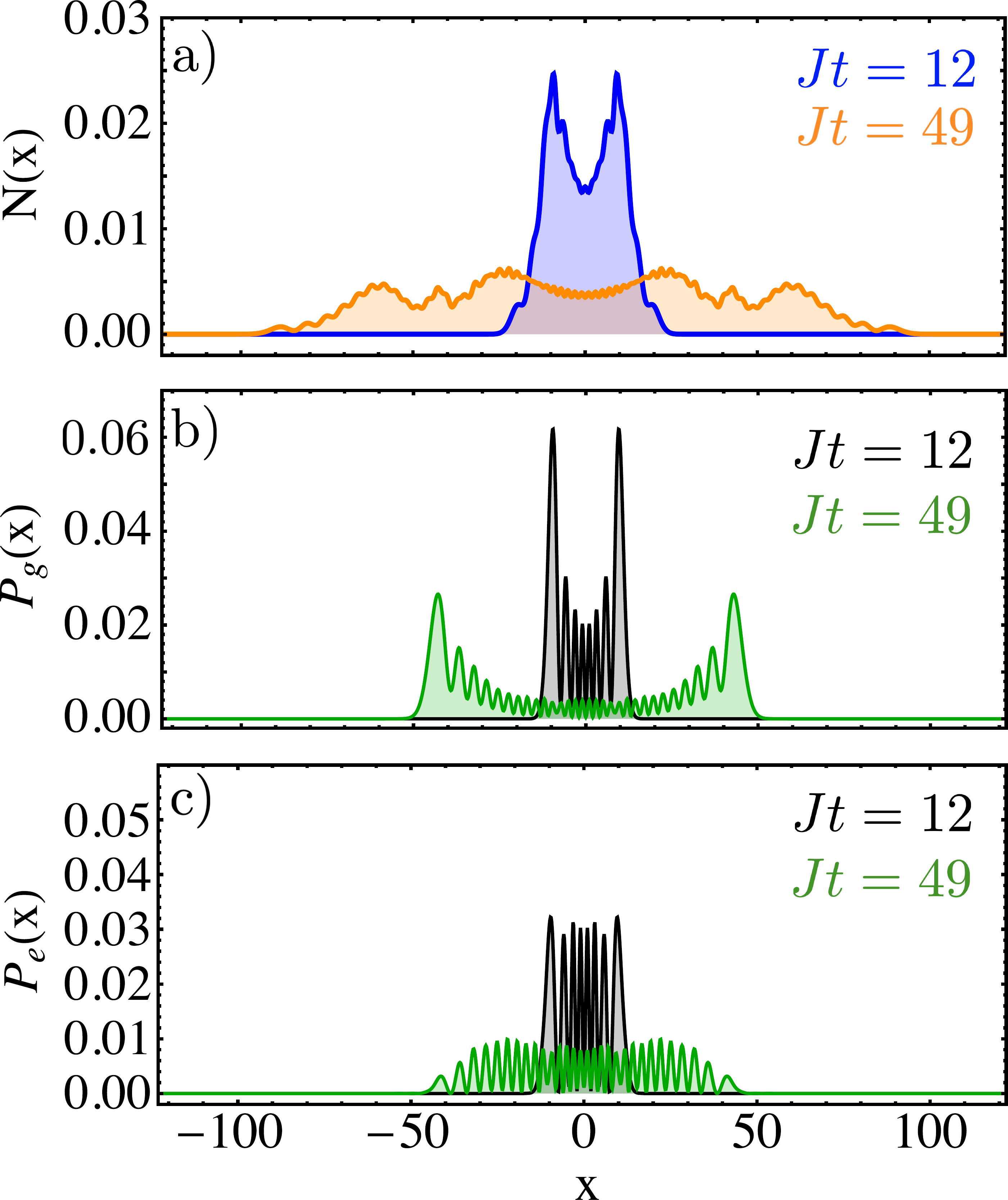}
\caption{Photonic (a), ground-state (b), and excited-state occupation (c) as a function of position, Eqs.~\eqref{Nxdef},\eqref{Pg}, and \eqref{Pe} respectively, for  $\Delta=-2.1J$, $J'=0.5J$, $\Omega=0.2J$, and two instants of time, namely $Jt=12$ (blue/black lines) and $Jt=49$ (orange/green lines).}
\label{fig:auxDeltam2p1}
\end{figure}

In this Appendix we show additional figures that complement our discussion in the main text. First, in Fig.~\ref{fig:v_appendix}, we extend the results of Fig.~\ref{fig:v} to the anti-parallel scattering case $\text{sign}[p_i]\ne \text{sign}[k_i]$. Our analysis of the scattering processes in Sec. \ref{SubsecProcesses} is general and thus applies also to this situation. Note that we can also label the two processes ``transmission'' (blue lines, corresponding to $ v^{\rm ph}_{p_{f,1}} < 0$ and $\vert v^{\rm ph}_{p_{f,1}}\vert > v^{\rm qb}_{k_{f,1}}$) and ``reflection'' (red lines, corresponding to $ v^{\rm ph}_{p_{f,2}} > 0$ and $ v^{\rm ph}_{p_{f,2}}> v^{\rm qb}_{k_{f,2}}$).

In Fig. \ref{fig:emission_K_appendix}(a) we show the photonic occupation at each momenta $p$ for spontaneous emission of an initially excited qubit with momentum $K=\pi/3$ (compare with Fig.~\ref{fig:emission_K}b). The profile at short and long times is similar to the corresponding profile at $K=0$, but with the peaks shifted to the corresponding values of $p_\pm$. The height asymmetry between the long-time peaks (blue lines) is only apparent, and disappears at asymptotically long times (one can indeed check that the area below both peaks is the same). In Fig.~\ref{fig:emission_K_appendix}(b) we show the effective photonic band for $K=0$ (c.f. Fig.~\ref{fig:emission_K}b) and for $K=\pi/3$. In both cases, the two asymptotic values of the emitted photon momenta, $p_\pm$ (dashed lines), lie at the same distance to the band minimum. This symmetry results in equal density of states at $p=p_+$ and $p=p_-$, and thus in non-directional emission as discussed in the main text.

In Figs. \ref{fig:auxDelta3} and \ref{fig:auxDeltam2p1} we show the position-dependent occupations of the waveguide, the qubit ground state, and the qubit excited state for two cases where emission is $K-$selective, namely $\Delta=3J$ (red line in Fig.~\ref{fig:xi_up}) and $\Delta = -2.1J$ (red line in Fig.~\ref{fig:xi_low}) respectively. In the case $\Delta=3J$, the emission is suppressed at values of $K$ near the edge of the Brillouin zone (see Fig.~\ref{fig:xi_up}). This results in a suppression of the fast propagating components of the emitted wavepackets in panels \ref{fig:auxDelta3}a) and  \ref{fig:auxDelta3}b), with respect to the case $\Delta=0$ (compare with Figs.~\ref{fig:Nx} and \ref{fig:ExGx}). For the same reason, the excited-state probability distribution is peaked at its wavefront (large momenta). Conversely, for $\Delta=-2.1J$ (Fig.~\ref{fig:auxDeltam2p1}), only the components with wavevectors near $\sim \pi/2$ are emitted (see Fig.~\ref{fig:xi_low}). This results in
emitted photonic and ground-state qubit wavepackets with a higher degree of coherence, as part of the momenta have been ``filtered out'' (compare panels \ref{fig:auxDeltam2p1}a and \ref{fig:auxDeltam2p1}b with the case $\Delta=0$ in Figs.~\ref{fig:Nx} and \ref{fig:ExGx}). Regarding the excited-state wavepacket in both Figs. \ref{fig:auxDelta3}c) and \ref{fig:auxDeltam2p1}c), its coherence  decreases with respect to the case $\Delta=0$, as the excited-state component of the total quantum state does not contain a single momentum $K$ anymore, but the whole range of momenta for which emission is suppressed. Finally, note that the total excited state probability, i.e. the integral of the curves in panels \ref{fig:auxDelta3}c) and \ref{fig:auxDeltam2p1}c), is larger than for the $\Delta=0$ case due to the partial inhibition of the spontaneous emission (see also Fig.~\ref{fig:pe_t}).

\bibliographystyle{apsrev4-1}
\bibliography{bib_movingqubit}

\end{document}